\documentclass[aps,physrev,twocolumn,superscriptaddress]{revtex4-2}

\usepackage[utf8]{inputenc} 
\usepackage[T1]{fontenc}    
\usepackage{hyperref}       
\usepackage{url}            
\usepackage{booktabs}       
\usepackage{amsfonts}       
\usepackage{nicefrac}       
\usepackage{microtype}      
\usepackage{xcolor}         

\usepackage{amsmath}
\usepackage{graphicx}
\usepackage{wrapfig}
\usepackage{subfig}
\usepackage{soul}
\usepackage{caption}

\newcommand{\X}{\mathcal{X}}
\newcommand{\M}{\mathcal{M}}
\newcommand{\LL}{\mathbf{L}}
\newcommand{\rrr}{\mathbb{R}}

\NewDocumentCommand{\R}{o}{\IfValueTF{#1}{\mathbb{R}^{#1}}{\mathbb{R}}}
\NewDocumentCommand{\hem}{}{hemagglutinin }
\NewDocumentCommand{\ig}{}{IgG }
\NewDocumentCommand{\snr}{}{$\mathrm{SNR}$ }
\NewDocumentCommand{\snrt}{}{$\widetilde{\mathrm{SNR}}$ }
\NewDocumentCommand{\dist}{O{k}}{\mathrm{dist}_{#1}}

\begin{document}


\title{Cryo-EM images are intrinsically low dimensional}

\author{Luke Evans}\thanks{Equal Contribution}
\affiliation{Center for Computational Mathematics, Flatiron Institute, New York, NY, USA}
\author{Octavian-Vlad Murad}
\thanks{Equal Contribution}
\affiliation{Department of Statistics, University of Washington, Seattle, Washington, USA}
\author{Lars Dingeldein}
\affiliation{Frankfurt Institute For Advanced Study, Frankfurt, Hesse, Germany}
\author{Pilar Cossio}
\affiliation{Center for Computational Mathematics, Flatiron Institute, New York, NY, USA}
\affiliation{Center for Computational Biology, Flatiron Institute, New York, NY, USA}
\author{Roberto Covino}
\affiliation{Institute of Computer Science, Goethe University Frankfurt, Frankfurt, Hesse, Germany}
\affiliation{Frankfurt Institute For Advanced Study, Frankfurt, Hesse, Germany}
\author{Marina Meila}
\affiliation{Department of Statistics, University of Washington, Seattle, Washington, USA}



\date{\today}

\begin{abstract}
    { Simulation-based inference provides a powerful framework for cryo-electron microscopy, employing neural networks in methods like CryoSBI to infer biomolecular conformations via learned latent representations. This latent space represents a rich opportunity, encoding valuable information about the physical system and the inference process. Harnessing this potential hinges on understanding the underlying geometric structure of these representations. We investigate this structure by applying manifold learning techniques to CryoSBI representations of { a simulated benchmark dataset, and both simulated and experimental images of} hemagglutinin. We reveal that these high-dimensional data inherently populate low-dimensional, smooth manifolds, with simulated data effectively covering the experimental counterpart. By characterizing the manifold's geometry using Diffusion Maps and identifying its principal axes of variation via coordinate interpretation methods, we establish a direct link between the latent structure and key physical parameters. Discovering this intrinsic low-dimensionality and interpretable geometric organization not only validates the CryoSBI approach but enables us to learn more from the data structure and provides opportunities for improving future inference strategies by exploiting this revealed manifold geometry.}
\end{abstract}



\maketitle

\section{Introduction}

Cryogenic electron-microscopy (cryo-EM) is a structural biology technique for imaging individual biomolecules at atomic resolution.
In a cryo-EM experiment, a biomolecular sample is imaged with a transmission electron microscope, and the resulting data is processed to yield a large dataset of unlabeled 2D images with one molecule per image (particles).   
Reconstruction algorithms \cite{nogales2015cryo} can estimate the 3D structure of the biomolecule from the 2D particles. In many cases, biomolecules coexist in different conformational states in the sample. 

Machine learning methods, including diffusion maps \cite{Dashti2014} and deep-generative models \cite{zhong2021cryodrgn, chen2021deep, punjani20213dflex}, have become central in cryo-EM for reconstructing heterogeneous conformations of biomolecules \cite{donnat2022deep,tang2023conformational}.  These methods project the high-dimensional conformational space on to a low-dimensional latent representation, but these latent spaces lack interpretability \cite{lederman2023manifold}. { Some of these challenges can be addressed by incorporating physical interpretability during training \cite{klindt2024towards} or by using physics-based priors in the model \cite{shi2024deep}. Additionally, analyses based solely on simulated data have also provided valuable insights \cite{jeon2024cryobench}}. However, extracting physical { and geometrical} information from the featurized images remains challenging due to non-linear feature mapping, low signal-to-noise ratio (SNR) and uncertainty in pose assignment, which can be confused with conformational changes.

\begin{figure*}
    \centering
\includegraphics[width=\linewidth]{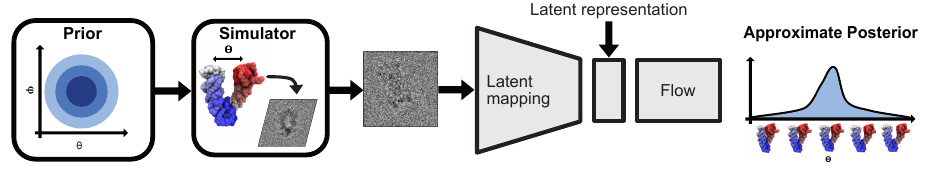}
    \caption{Schematic workflow for learning the surrogate posterior with cryoSBI. Parameter samples are drawn from the prior to simulate synthetic cryo-EM images. These images are then used to approximate the posterior by jointly training a summary network and a normalizing flow.} 
    \label{fig:explain}
\end{figure*}

Recent simulation-based techniques from integrative structural biology~\cite{rout2019principles} and probabilistic machine learning~\cite{murphy2022probabilistic} hold great promise for analyzing cryo-EM data.
CryoSBI~\cite{dingeldein2025amortized} { is an emerging paradigm using} simulation-based inference~\cite{cranmer2020frontier,papamakarios2016fast} (SBI) to infer conformations and  uncertainties from cryo-EM particles by training { neural networks producing a} 
latent representation
and { a density estimator} with simulated cryo-EM experiments {(Figure~\ref{fig:explain}}). The trained networks can be quickly evaluated on large experimental particle datasets.  Because the training is only done with simulated data, a key feature of cryoSBI is that it enables linking of physical properties of the molecules and the experiment to experimental data.

{ Supported by preliminary evidence~\cite{dingeldein2025amortized}}, we hypothesize that the  representations learned by the neural  network are near low dimensional manifolds inside the latent space. 
The objective of this work is to study the geometry of the data using manifold learning techniques \cite{coifman2005geometric,chen2019selecting, koelle2024consistency, meilua2024manifold}. First, we will seek to ascertain whether the learned representations correspond to well-behaved low-dimensional manifolds, and second, whether these are parameterized by generative variables important in predicting the posterior over the conformation. 
Our analysis quantitatively validates the latent space of cryoSBI and leads to a general computational workflow both for interpreting latent spaces of cryo-EM heterogeneity analysis methods and more broadly for learned summary statistics in simulation-based inference.

\section{CryoSBI and Latent Spaces}

CryoSBI~\cite{dingeldein2025amortized} is a new method to quantify the probability that a given image $I$ depicts a molecular conformation $\theta$. We assume to have a set of structures, e.g. from molecular simulations or { AI-methods \cite{monteiro2024high}}, which we expect to find in the sample. For simplicity, we also assume that $\theta$ is a one-dimensional parameter, and we aim to infer the conformation $\theta$ of the molecule observed in the image, i.e., compute the Bayesian posterior $p(\theta | I)$. The posterior quantifies how compatible $\theta$ is with the observed image $I$. 

To model the image formation process, one must consider experimental details such as microscope aberration, noise, and random orientation of the molecule. To simulate a cryo-EM image, one samples conformations from the prior $\theta_i \sim p(\theta)$, and imaging parameters from $\phi_i \sim p(\phi)$ and then generates a synthetic image $I_i \sim p(I| \theta_i, \phi_i)$ using a forward model of the imaging process (Appendix \ref{imageformation}), accumulating a data set of simulated images and ground truth parameters $\mathcal{D} = \{\theta_i, \phi_i, I_i\}_{i=1}^N$. The nuisance parameter vector $\phi_i$ includes random orientations, a wide range of defocus values, center translations, and SNRs. 

\subsection{Feature Latent Representation and Neural Posterior Estimation.}
\label{subsec:feature}
CryoSBI builds on the Neural Posterior Estimation framework~\cite{lueckmann2021benchmarking, zammit2024neural}, jointly training a latent representation network $S_{\psi}(\cdot)$ to extract summary statistics and a normalizing flow $q_{\varphi}(\cdot)$ as surrogate model of the posterior $q_{\varphi}(\theta|S_{\psi}(I)) \approx p(\theta | I)$. This is done by maximizing the average log-likelihood $\mathcal{L}(\varphi, \psi) = \frac{1}{N} \sum_{i=1}^{N} \log q_\varphi(\theta_i|S_\psi(I_i))$ of the posterior probability under the training samples $\mathcal{D}$ (Appendix \ref{sbinetworks}).
In principle, after training, $S_{\psi}$ should \textit{i}) compress images to predict the relevant features and \textit{ii}) enable efficient comparison of simulated images to `nearby' experimental images. 
For example, the latent representation should distinguish images due to conformation, SNR and projection direction, as these are the primary experimental factors determining how precisely we can estimate a molecular configuration from a single image. {Distinguishing these factors is another step towards indicating physical properties of the molecule, such as symmetries affecting the pose or conformation estimates.}
In practice, while the feature representation for { cryoSBI~\cite{dingeldein2025amortized} -- and more generally for Neural Posterior Estimation} -- offers powerful inference capabilities, it is not immediately interpretable, making it challenging to check for model misspecification~\cite{schmitt2024detecting}. 

\subsection{{ Datasets.}}
\label{subsec:datasets}
{ 
We consider latent representations from two different biomolecular systems: the human immunoglobulin G antibody (IgG) and influenza hemagglutinin. The IgG dataset was recently utilized as an example of large conformational changes (the ``IgG-1D'' dataset) in the cryo-EM heterogeneity benchmark CryoBench~\cite{jeon2024cryobench}. We applied cryoSBI to this benchmark system and analyzed the latent representations of simulated images derived from the IgG atomic models.
CryoSBI training was performed as in  \cite{dingeldein2025amortized} using cryo-EM simulations by sampling the priors with ranges specified in (Appendix C). After training, we created a simulated dataset $\mathcal{D}_s$ consisting of $N_s = 100,000$ feature vectors with $i$-th datapoint $x_i = S_{\psi}(I_i) \subseteq \mathbb{R}^{256},$
nuisance parameters $\phi_i,$ ground-truth conformation parameter $\theta_i$, posterior mean $\hat{\theta}_i$ and width $\sigma_i$ 
of the posterior $q_{\varphi}(\cdot |  x_i)$,
so that $\mathcal{D}_s = \{x_i, \hat{\theta}_i, \sigma_i, \theta_i, \phi_i\}_{i=1}^{N_s}.$
We denote the representations learned by $S_\psi$, $\mathcal{X}_s = \{x_i\}_{i=1}^{N_s}.$
}

The hemaglutinin data corresponds to the hemagglutinin dataset considered in ref.~\cite{dingeldein2025amortized}; it consists of the CryoSBI latent representations of the simulated and experimental images. CryoSBI training was performed as in  \cite{dingeldein2025amortized} by sampling the priors (Appendix \ref{priors}) and simulating from them. Likewise, we created a simulated dataset $\mathcal{D}_s$ consisting of $N_s = 100,000$ feature vectors with $i$-th datapoint $x_i = S_{\psi}(I_i) \subseteq \mathbb{R}^{256},$
nuisance parameters $\phi_i,$ ground-truth conformation parameter $\theta_i$, posterior mean $\hat{\theta}_i$ and width $\sigma_i$ 
of the posterior $q_{\varphi}(\cdot |  x_i)$,
so that $\mathcal{D}_s = \{x_i, \hat{\theta}_i, \sigma_i, \theta_i, \phi_i\}_{i=1}^{N_s}.$
The experimental dataset $\mathcal{D}_e$ consists of $N_e = 271558$ tuples $\mathcal{D}_e = \{\tilde{x}_i, \hat{\theta}_i, \sigma_i\}_{i=1}^{N_e}$ with $\tilde{x}_i = S_{\psi}(\tilde{I}_i)$, for whitened single particle-images $\{\tilde{I}_i\}_{i=1}^{N_e}$ from EMPIAR 10532~\cite{tan2020through}, where $\hat{\theta}_i, \sigma_i$ are the inferred posterior parameters (note that
the experimental images have no ground truth $\theta$ or $\phi$). 
We denote the representations learned by $S_\psi$, $\mathcal{X}_s = \{x_i\}_{i=1}^{N_s}$ and $\mathcal{X}_e = \{\tilde{x}_i\}_{i=1}^{N_e}$.


{\section{Manifold analysis pipeline}}
\label{sec:manifold_analysis}

{ To study the representations learned by the feature mapping $S_{\psi},$ we apply a framework of manifold learning techniques to the datasets $\mathcal{X}_s$ and $\mathcal{X}_e$ (Appendix~\ref{sup_sec:framework}). We proceed by a pipeline of: pre-processing the data, estimating the intrinsic dimensionality, and then embedding to a lower dimensional manifold. 
}

\subsection{Data pre-processing.}
{ 
The data pre-processing consists of subsampling, removing outliers, and resampling the data to avoid large variation in density.
The data subsampling and outlier removal ensure both computational scalability and statistical robustness in our methods, and the re-sampling encourages a more uniform distribution of data, which is known to help de-bias both dimension estimation and manifold learning techniques~\cite{meilua2024manifold}. Our full process
is described in detail in Appendix \ref{sup_sec:pre-processing}. 
}
\subsection{{ Dimensionality estimation}.}
{ Intrinsic dimension estimation is challenging when the data contains noise in the   high dimensional ambient space. To ensure robustness of our findings, we employ four different estimators. The first returns a global dimension estimate $\hat{d},$  and the latter three return a local dimension estimate $\hat{d}(x)$ for any given data point $x;$ in this second case, the estimate $\hat{d}$ is the mode of the $\hat{d}(x)$ distribution. 

The four algorithms we use are:
\begin{itemize}
    \item \textit{Correlation Dimension}~\citep{grassberger1983slope}:  
    The number of neighbors within a radius $r$ of a datapoint $x$ satisfies $n_r(x) \propto r^d$, implying $\log n_r(x) \propto d \cdot \log r$. For a range of radii $R=\{r_1,r_2,\ldots r_m\}$, we compute the average number of neighbors $\overline{n_r}$ 
    and fit a linear model to $\{(\log r, \log \overline{n_r})\}_{r \in R}$. The slope yields a single global estimate $\hat{d}$ (Figure~\ref{fig:d_estim}a).

    \item \textit{Eigengap Method}~\citep{chen2013eigengap}:  
    For each $x$, we compute a weighted local covariance matrix and obtain its truncated spectrum 
    $\lambda_1 \geq \dots \geq \lambda_D$ for some conservatively large $D > d$. 
     Then $\hat{d}(x)$ is given by  the largest gap between consecutive eigenvalues. This method is the only one of the four designed to take into account the presence of noise.  
    (Figure~\ref{fig:d_estim}b). 
    \item \textit{Doubling Dimension}~\citep{assouad1983doubling}:  
    By the same scaling principle as Correlation Dimension, we can instead compute the local statistic $\hat{d}_r(x) = \log_2\big(n_{2r}(x)/n_{r}(x)\big)$ for some fixed $r \in \R^+$(Appendix Figure~\ref{fig:dd_all}). These local estimates can be averaged over a range of radii $R$ to yield more robust point-wise predictions (Figure \ref{fig:d_estim}c).

    \item \textit{Levina–Bickel MLE}~\citep{levina2004MLEdim}:  
    This method infers $d$ from the top $k$-nearest neighbor distances as $\hat{d}_k(x) = \big[ \frac{1}{k-1} \sum_{j=1}^{k-1} \log (\dist[k](x)/\dist[j](x)) \big]^{-1}$ for some fixed integer $k$ (Appendix Figure~\ref{fig:lb_all}). These local estimates can be averaged over a range of numbers of neighbors $K$ to yield more robust point-wise predictions (Figure \ref{fig:d_estim}d).
\end{itemize}
In Appendix \ref{sup_sec:dim_estim} and Appendix \ref{sup_sec:cover} we provide details about the algorithms, their implementation in this work,   and practical guidelines for their use.
}

\subsection{{ Manifold embedding}.}

{ For embedding the high-dimensional datasets to lower dimensions, we 
apply Diffusion Maps~\citep{coifman2005geometric} to our pre-processed data. To refine our initial embedding, we remove redundant embedding directions by applying Independent Eigencoordinate Selection (IES)~\citep{chen2019selecting}. We then correct the distortion  introduced by the embedding algorithm to the original distances within the manifold, through the procedure known as Riemannian Relaxation~\cite{mcqueen2016relaxation}. Appendix~\ref{sup_sec:embed} details our embedding pipeline.
}
\section{Geometric Analysis of the CryoSBI Latent Space}
 Now, we proceed to study the shape of the data cloud $\X_e$, 
 determine the intrinsic dimensionality of the datasets, assess how well the simulated data covers the experimental space, and uncover the physical interpretation of the latent representations.

\subsection{Are the data low dimensional?}\label{subsec:dim_estim}

\begin{figure*}
    \includegraphics[width=0.8\linewidth]{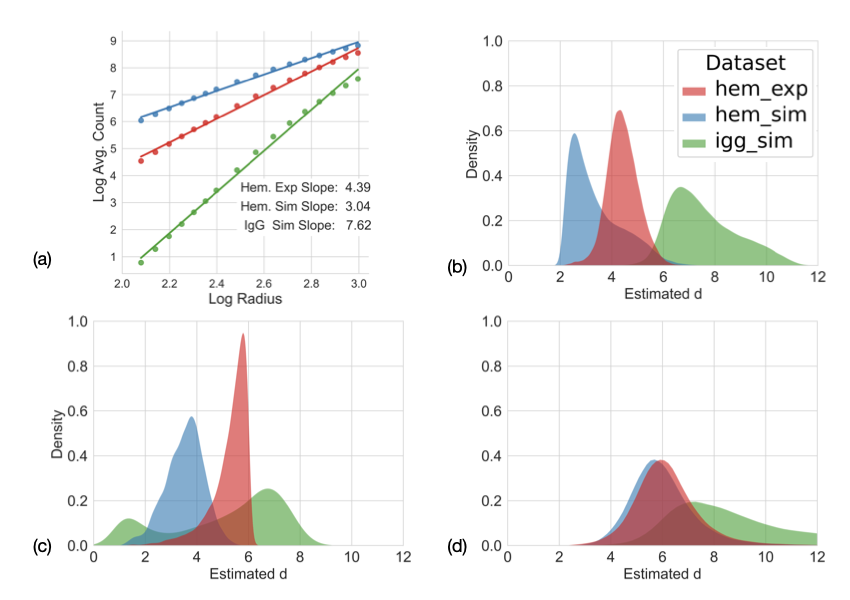}
    \caption{
    \raggedright
   { 
    Intrinsic dimensionality estimation results for the experimental (red) and synthetic (blue) \hem datasets, and the synthetic \ig dataset (green).  \newline
    (\textbf{a}) \textit{Correlation Dimension}: The slope of the fitted line yields a single global prediction $\hat{d}$ for each dataset. \newline
    (\textbf{b}) \textit{Doubling Dimension}: distribution of the point-wise estimates   $\hat{d}(x)$ averaged over a range of radii $R.$ \newline
    (\textbf{c}) \textit{Eigengap}: distribution of the point-wise intrinsic dimension estimates $\hat{d}(x)$. \newline
    (\textbf{d}) \textit{Levina–Bickel}: distribution of the point-wise estimates  $\hat{d}(x)$ averaged over a range of nearest neighbors $K$.
    }
    }
    \label{fig:d_estim}
\end{figure*}



{ We  first estimate the intrinsic dimension for the synthetic \ig dataset and the experimental and synthetic \hem datasets. Figure \ref{fig:d_estim} summarizes our main result:
} all { four} methods indicate that the { learned cryo-EM embeddings} have {\em low intrinsic dimension.}

{ Specifically, we find that $6 \leq d_s \leq 10$ for the synthetic \ig dataset and $2 \leq d_s \leq d_e \leq 6$ for the \hem datasets. 
The higher dimensionality for simulated IgG data than the simulated hemagglutinin is likely due to the larger conformational change in the IgG pdbs.
The slight difference between $d_s$ and $d_e$ for hemagglutinin likely reflects experimental noise and other imaging factors not captured by the simulated model. 
Taken together, these results strongly support the conclusion that the manifold assumption holds for our datasets.  
}

\subsection{Does the simulated hemagglutinin data cover the experimental data well ?}\label{subsec:cover}

For the amortized simulation-based-inference in CryoSBI, the simulator must be able to generate many relevant experimental realizations, so that a particular experiment can be accurately analyzed without retraining. Our dimensionality results above indicate that the simulated and experimental { hemagglutinin} data lie on low dimensional manifolds, but do not inform whether the manifolds are close to each other or if the simulated data covers the experimental. In other words, if the experimental { hemagglutinin} data are in the distribution of simulated ones. 

 We investigate the covering by density estimation of both { hemagglutinin} datasets.  For this, we first estimate the data densities $p_e$ and $p_s$ in $\rrr^{256}$ by kernel density estimators (KDE)~\citep{silverman2018density} $\hat{p}_e$ and $\hat{p}_{s}$. 

The bandwidths $h_e = 0.31$ and $h_s=0.48$ are obtained by cross-validation, { as detailed in Appendix~\ref{sup_sec:cover}.}
 While it is known that KDE is poor in high dimensions, the method is {\em adaptive}, meaning that it will work when the intrinsic dimension is low, as in this case.  We use samples of size 17000 for fitting $\hat{p}_e$ and $\hat{p}_{s}$. We do not expect $p_e$ to equal $p_s$, but we would like to confirm that $p_s$ is predictive of the experimental data. Thus, on two held out
datasets $\mathcal{X}^{test}_{e}$ and $\mathcal{X}^{test}_{s}$, with $|\mathcal{X}^{test}_{s}| =
|\mathcal{X}^{test}_{e}| = n^{test}= 3000$, we calculate the negative log-likelihoods (i.e., cross-entropies) $-\frac{1}{n^{test}}\log \hat{p}_{m}(\X^{test}_{m})$ for $m \in \{e, s\}$(in Table \ref{tab:kde_results}) and the estimated Kullback-Leibler divergences $D_{KL}(\hat{p}_e || \hat{p}_{s})=89.0$, $D_{KL}(\hat{p}_{s} || \hat{p}_{e})=2417.9$. These show
that the simulated data can predict the experimental data well; meanwhile, the experimental data does not completely cover the simulated data. For further analysis, we retain in $\X_s$ only the samples that are near the experimental data. The hypothesis that we can 
infer what generative parameters best describe the experimental data, is so far supported since we can, for most experimental $\tilde{x} \in \mathcal{X}_e$, find enough near-by
synthetic $x \in \mathcal{X}_s$ to perform this prediction in a robust
manner. 

\begin{table}[b]
    \begin{tabular}{|c|c|c|}
    	\hline
            \multicolumn{3}{|c|}{$-\frac{1}{n}\log \hat{p}_{m}(\X_m^{test})$} \\
            \hline
    	& $\X^{test}_e$  & $\X^{test}_s$ \\
    	\hline
    	$\hat{p}_e$ & 86.6 & 2604.3 \\
    	$\hat{p}_s$ & 175.6 & 186.4 \\
    	\hline
    \end{tabular}
    \caption{Test data negative log-likelihoods under $\hat{p}_e$ and $\hat{p}_s$.}
    \label{tab:kde_results}
\end{table}

\subsection{Modeling the low dimensional cryo-EM images manifold.} \label{subsec:embed}

\begin{figure*}
  \begin{minipage}{.45\textwidth}
    \centering
    \subfloat[$\Phi_e$ colored by $\tilde{\theta}$ (left) and $\hat{\theta}$ (right).]{
      \includegraphics[width=0.45\textwidth]{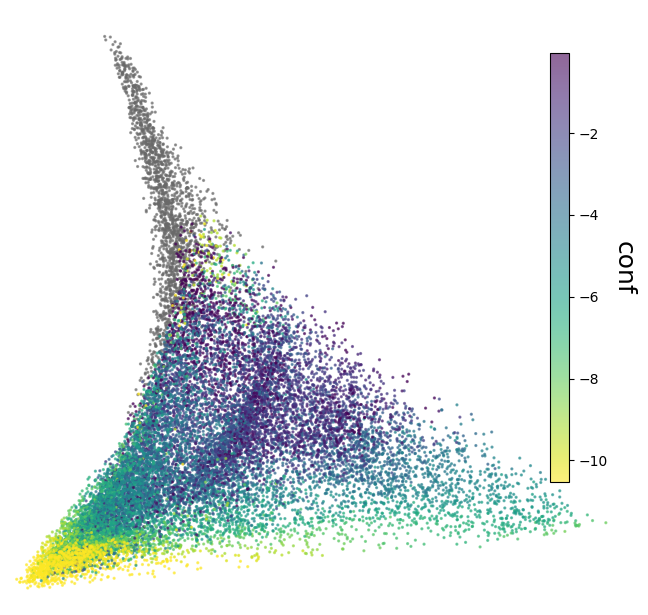}
      \includegraphics[width=0.45\textwidth]{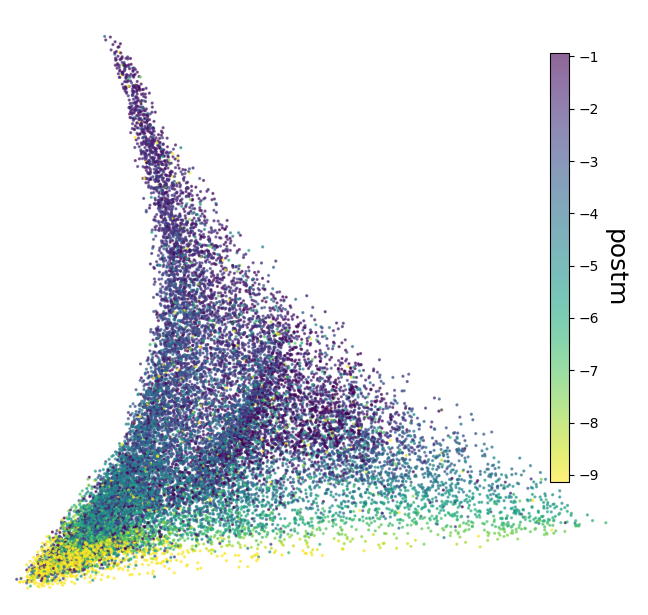}
    }
  \end{minipage}%
  \hfill
  \begin{minipage}{.51\textwidth}
    \centering
    \subfloat[$\Phi_s$ colored by $\theta$ (left) and $\hat{\theta}$ (right).]{
      \includegraphics[width=0.5\textwidth]{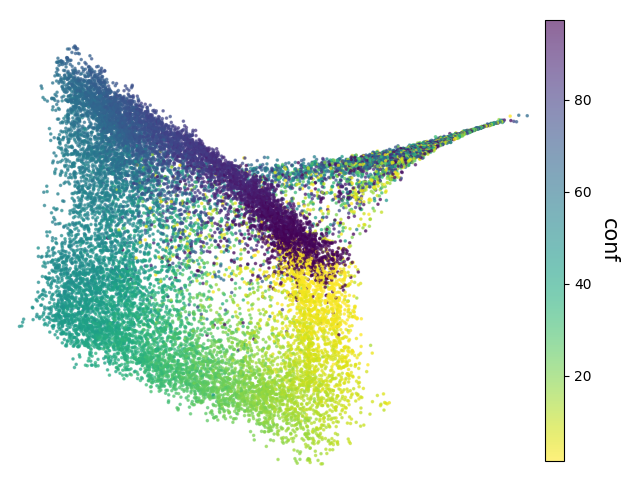}
      \includegraphics[width=0.5\textwidth]{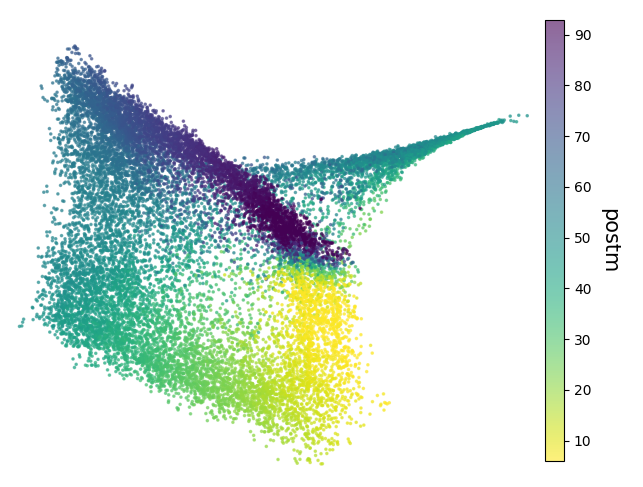}
    }
  \end{minipage}

  \vspace{.02\textwidth}

  \begin{minipage}{.45\textwidth}
    \centering
    \subfloat[$\Phi_e$ colored by $\widetilde{\mathrm{SNR}}$ (left) and posterior width $\sigma$ (right), with example particles corresponding to indicated regions.]{
      \includegraphics[width=0.425\textwidth]{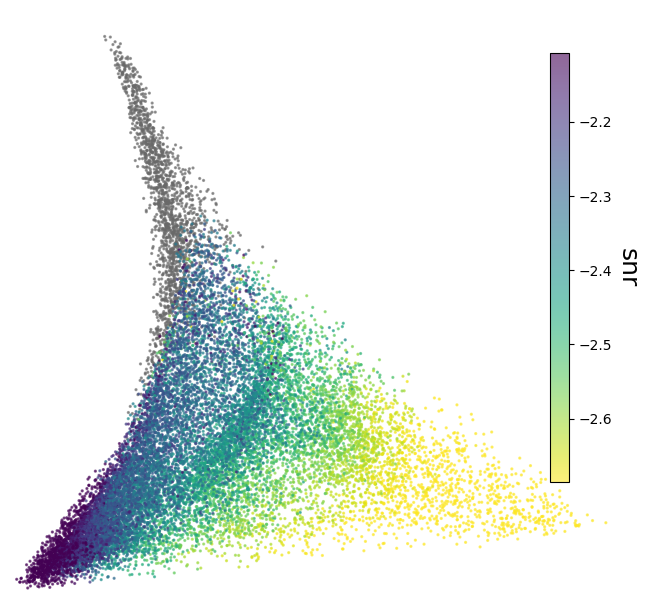}
      \includegraphics[width=0.425\textwidth]{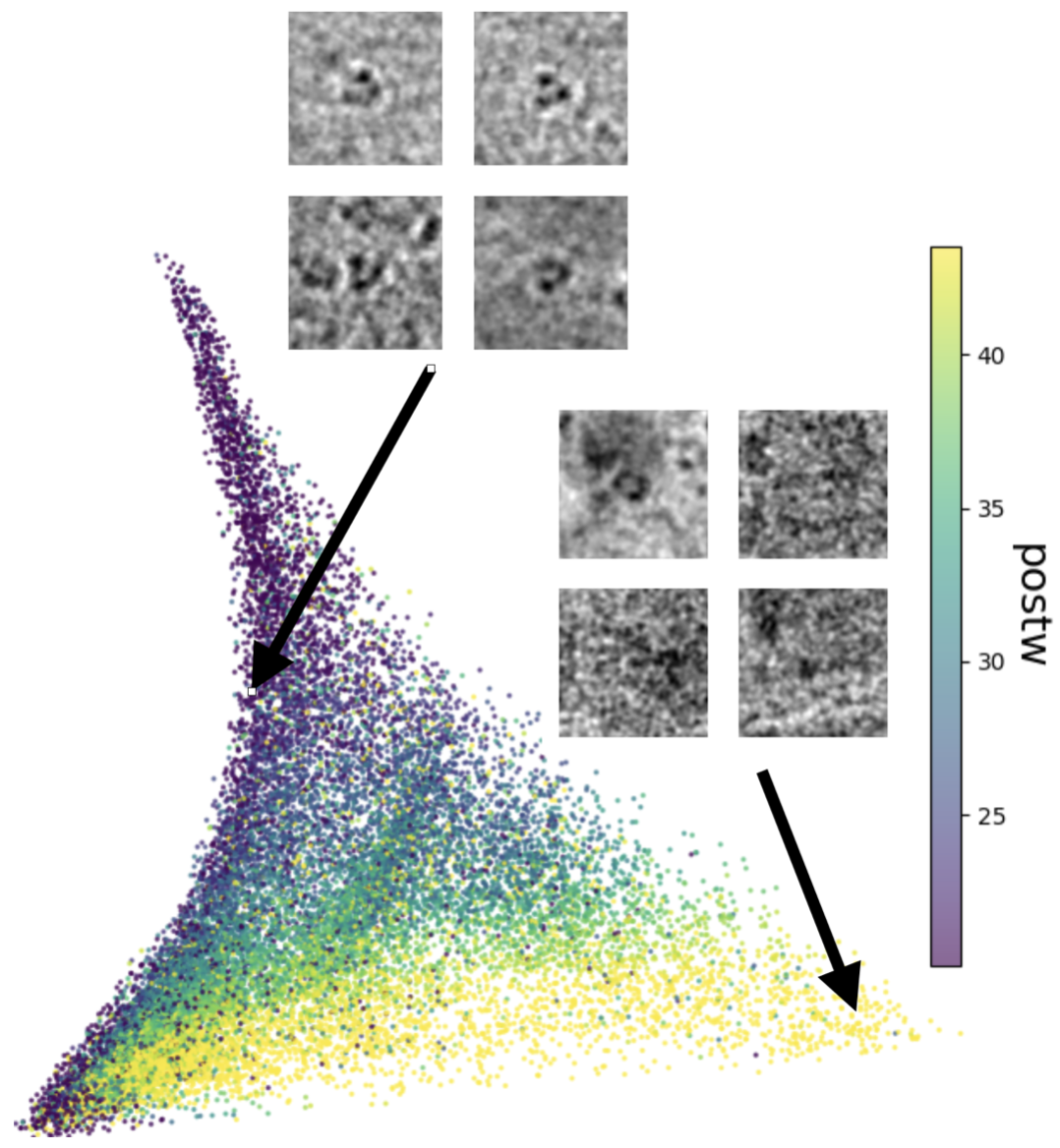}
    }
  \end{minipage}%
  \hfill
  \begin{minipage}{.51\textwidth}
    \centering
    \subfloat[$\Phi_s$ colored by $\mathrm{SNR}$ (left) and posterior width $\sigma$ (right).]{
      \includegraphics[width=0.5\textwidth]{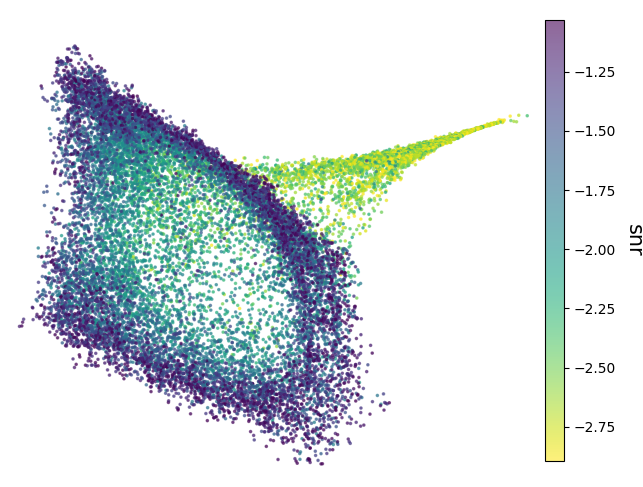}
      \includegraphics[width=0.5\textwidth]{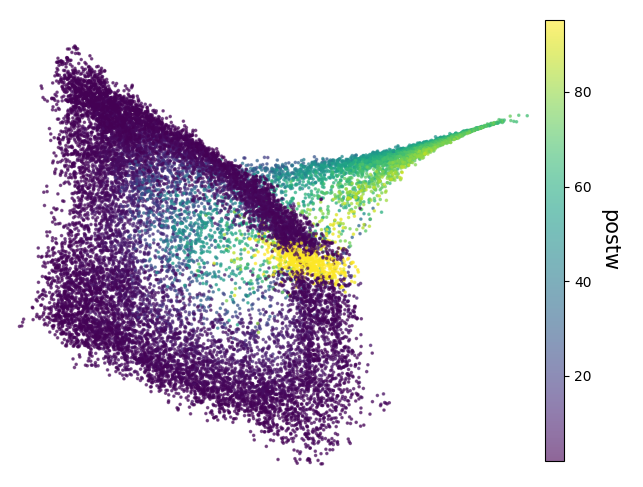}
    }
  \end{minipage}

  \caption{
  { Diffusion Maps embedding $\Phi_e$ learned from the experimental \hem dataset \textbf{(a), (c))} and $\Phi_s$ learned from the simulated \ig dataset \textbf{((b), (d))}. We display the lowest frequency $d=3$ coordinates selected by IES and rotate the plots to best display the embedding. Importantly, the neural network predicted conformation $\hat{\theta}$ and uncertainty $\sigma$ vary smoothly over the manifolds and agree with the true generative conformation $\theta$ and $\mathrm{SNR}$ over the embedded points. In the absence of the true generative conformation $\theta$ and $\mathrm{SNR}$ for the experimental data, we use $\widetilde{\theta}$ and $\widetilde{\mathrm{SNR}}$ to color $\Phi_e$, values which we infer via kernel interpolation (see Appendix \ref{sup_sec:interpolation}) from the simulated \hem dataset. Plots \textbf{(c)} and \textbf{(d)} have opposing color scales to emphasize regions where (high/low)$\mathrm{SNR}$ correspond to (low/high) posterior width.}
  }
  \label{fig:exp_igg_embeddings}
\end{figure*}

We use a suite of manifold learning techniques ~\cite{coifman2005geometric,
  perraultjoncas2014consistency, chen2019selecting,
  mcqueen2016relaxation} to map the neural representations
$\mathcal{X}_e \subseteq \mathbb{R}^{256}$ { of the experimental hemagglutinin data} down to a much
lower dimensional embedding $\Phi_e$, which we here interpret geometrically and in the following section from the physical point of view, in relation to the simulated { hemagglutinin} data $\X_s$.

We use Diffusion Maps
\cite{coifman2005geometric} with a kernel width parameter $\epsilon$
selected by the method of ref. \cite{perraultjoncas2014consistency} to
compute the low-dimensional embedding $\Phi_e\in\rrr^d$ of $\X_e$ ({ Figure~\ref{fig:exp_igg_embeddings}}); similarly we compute $\Phi_s$ for the filtered $\X_s$ data {, for both the simulated IgG data and simulated hemagglutinin (Figure~\ref{fig:exp_igg_embeddings}, Appendix Figure \ref{fig:sim_embeddings} respectively). 
The embeddings indicate that conformation and SNR are primary factors influencing the structure of the data manifolds, and they vary in regions that closely align with the posterior mean and posterior width, respectively.
See Appendix~\ref{sup_sec:embed} for further practical details on the pre-embedding filtering and parameter selection, and Appendix Figures~\ref{fig:igg_embedding},\ref{fig:exp_embeddings} for further embeddings.}  

The Diffusion Maps embedding is based on the eigendecomposition of the Laplacian matrix $\LL$ \cite{coifman2005geometric}, and in a first stage we compute it up to the $m$'th non-zero eigenvalue, for $m=20$, and denote these coordinates with $\Phi_{1:m}\in\rrr^n$, with $n=|\X_e|$. The analysis of the principal eigenvalues of $\LL$, which are slowly growing and well above 0 (Appendix Figure \ref{fig:hem_eigvals}), indicates that the manifold $\M_e$ is connected, and likewise for the IgG manifold (Appendix Figure~\ref{fig:igg_eigvals}). That is, there are no isolated clusters and no outliers for the postprocessed data. 
{ However, this analysis does not rule out the presence of clusters as high-density regions, which could occur from data that is not pre-processed. { As a visual example of the effect of pre-processing (Appendix~\ref{sup_sec:pre-processing}), 
 Appendix Figure~\ref{fig:exp_density} maps a sample from the original probability density $p_e$ into $\M_e$.}}
 
Next, we perform IES \cite{chen2019selecting} to select $d=3$ independent and low-frequency coordinates from  $\Phi_{1:20}$. We use these coordinates, denoted $\Phi_e$, to visualize and interpret the experimental data { (see Appendix~\ref{sup_sec:embed} for more details)}.
As
shown previously, $d=3$ is likely close to
the true intrinsic dimension of $\mathcal{M}_s$ and $\mathcal{M}_e$,
meaning we can expect to capture most of the  relevant structure of the
experimental data by analysing these $\Phi_e$ coordinates. We apply Riemannian
Relaxation \cite{mcqueen2016relaxation} to $\Phi_e$ to make it closer to 
being isometric to $\mathcal{X}_e$ { for hemagglutinin}. The resulting embedding is shown in Figure \ref{fig:exp_embeddings}. We perform
similar steps with the simulated data $\X_s$ { for hemagglutinin} (Appendix \ref{sup_sec:embed}). 
\subsection{Physical interpretation of the experimental data manifold} \label{subsec:valid}
 \label{subsec:tslasso}
In the absence of ground truth generative parameters for the experimental
{ hemagglutinin} data, we have to find alternative ways to determine whether $S_\psi$ is a good predictor for the
true conformational parameter $\theta$, and the noise level, an
important nuisance parameter. While this can be done with a manually labeled test set, we focus on indirect geometric methods that don't require scientific labeling. We first use a statistical method, TSLasso \cite{koelle2024consistency} to interpret the embedding $\Phi_e$. Afterwards, we support its results and expand the analysis with visualizations. TSLasso searches for the optimal interpretation of an embedding in a {\em
  dictionary} $\mathcal{F} = \{f_k: \mathcal{M}_e \to \mathbb{R}, k={1:p}
\}$ of (smooth) potential coordinate functions on $\mathcal{M}_e$. Here, each $f_k \in
\mathcal{F}$ represents one of the simulation parameters (the
conformation $\theta$ or one of the nuisance parameters in $\phi$), hence $|\mathcal{F}| = 10 = p$.
TSLasso recovers a subset $f_S$
of $\mathcal{F}$ which parametrizes $\mathcal{M}_e$, by selecting $d$ functions whose gradients "most economically" span the tangent spaces of the manifold at a sample of the data. Since the functions $f_k$ are unknown on the experimental data, we infer them by interpolation (Appendix \ref{sup_sec:interpolation}), obtaining $\tilde{\theta}$ and
$\tilde{\phi}$ for the experimental data. We also estimate the gradients $\nabla f_k$ ({ Appendix \ref{sup_sec:grad_estim}}).  TSLasso is run 20 times using random subsets of 500 data points. We find that $f_S$ almost always consists of conformation $\theta$, SNR, and one of the rotation coordinates in $\phi/\tilde{\phi}$ (albeit not
always the same one). The full results are presented in Appendix \ref{sup_sec:tslasso}. For completeness, we apply the same algorithm to the simulated data {{ of both hemagglutinin and IgG}. Our results show that this combination of functions parametrizes both $\mathcal{M}_s$ and $\mathcal{M}_e$.
We have confirmed statistically, without any visualization, that the
two parameters $\theta$ and SNR inferred from nearby simulated data,
vary smoothly along the experimental data manifold $\M_e$ (as well as along
$\M_s$), therefore, supporting the neural network predictions for
$\X_e$. The visualizations are shown in { Figure \ref{fig:exp_igg_embeddings} (a) and (b)}.

\section{Discussion}

In summary, our study of the latent embedding representations of { IgG and} hemagglutinin cryo-EM data from cryoSBI, has revealed that these live near a well-behaved low dimensional manifold in $\rrr^{256}$ space. { The difference in intrinsic dimensionality of the simulated data manifolds, from hemagglutinin to the more heterogeneous IgG, indicate that conformational variability affects dimensinality of the manifold. For the hemagglutinin data,} the simulated { hemagglutinin} images cover (almost entirely) the experimental ones. Therefore, we can use the simulated data (on which we have full control) to interpret the experimental data in the latent space. Furthermore, we  have identified the physical and geometrical features that explain the different directions in the latent space.

We presented visualizations (e.g., by post-processed Diffusion Maps embedding) that accurately display the data shape by being almost isometric. We are also excited by the possibilities of replacing visual
analysis with quantitative measures, and principled algorithms in creating and validating low dimensional
models of cryo-EM data. Examples of such tasks include detecting the intrinsic dimensionality, interpreting the manifold by physical coordinates, measuring the smoothness of functions over the data manifold (not included here, but straightforward via the Laplacian operator), detecting if clusters exist, and measuring local distortion \citep{joncasRmetric}. 

From the methodological point of view, we present a pipeline for analyzing, exploring and visualizing high dimensional data presumably living near a smooth manifold. The pipeline components integrate state of the art geometric algorithms and theoretical results. However, we note that we do not propose to replace the trained neural network predictor with (a variant of) the methods presented here. Typically, dimension reduction methods do not outperform a neural network trained in supervised mode. What our method offers is
interpretability of the latent representations and a connection of the experimental data to the physical simulator.  

At the same time, we acknowledge that the data might not align perfectly with the manifold hypothesis. Our current understanding does not yet enable us to predict, comprehend, or control how finer-scale data structures— e.g., what we consider "noise"—affect geometric algorithms, which should be a matter of further investigation. 

{ We note that our work here investigates only one experimental dataset, and only a synthetic example of a large conformational change.
We showed that large conformational change appears to increase the intrinsic dimensionality, but other factors such as non-white noise and background effects could play a role.
We anticipate that our protocols will be robust when applied to data with non-uniform pose distributions or non-uniform conformational distributions.}

In future work, one may investigate if the settings of cryoSBI, such as the chosen dimension of the latent space, choice of priors on imaging parameters, and size of the training set affect the geometry, topology, and intrinsic dimension of the embedding.
}
{ \section*{Data Availability}
The code for the geometry analysis is available on GitHub at \url{https://github.com/ovmurad/cryosbi_manifolds}. The data are available on  
the Zenodo repository \url{https://zenodo.org/records/15733579}~\cite{evans_2025_15733579}.
}

\section*{Acknowledgments}
The Flatiron Institute is a division of the Simons Foundation.  L.D. and R.C. acknowledge the support of Goethe University Frankfurt, the Frankfurt Institute of Advanced Studies, the LOEWE Center for Multiscale Modelling in Life Sciences of the state of Hesse, the CRC 1507: Membrane-associated Protein Assemblies, Machineries, and Supercomplexes (P09), and the International Max Planck Research School on Cellular Biophysics.
L.D., and R.C. thank the Flatiron Institute, and M.M. gratefully acknowledges the DataShape Group at INRIA Saclay for hospitality while a portion this research was carried out.
This project was initiated out of discussions at the "Data Driven Materials Informatics"  long program at the Institute for Mathematical and Statistical Innovation (IMSI),  and was further motivated by the NeurIPS workshop "Machine Learning and Structural Biology". The authors would like to thank the organizers and IMSI for fostering this collaboration.




 

\bibliographystyle{unsrt}

\appendix
\section{Cryo-EM image formation forward model}
\label{imageformation}
We simulate cryo-EM particles from 3D molecular structures with the forward model of~\cite{seitz2019simulation, giraldo-barreto_BayesianApproach_2021}. The electron density $\rho(X)$ of a given structure $X$ is approximated as a Gaussian mixture model with centers on the positions of the $C_\alpha$ atoms, and standard deviations $\gamma.$ 
Then, we apply a rotation $R_q$ with quaternion $q$ and projection $P_z$ onto the $z-$axis to $\rho(X),$ then convolve with a point-spread function (PSF), which incorporates the microscope defocus and aberration. 
The PSF is more straightforward to apply in Fourier space, where the convolution becomes a point-wise multiplication with the Fourier transform of the point-spread function, known as the Contrast Transfer Function (CTF).
The CTF is defined as
$
   \mathrm{CTF}_{A,b, \Delta z}(s) = e^{-b s^2/2}\left[A \cos(\pi \Delta z \lambda_e s^2) - \sqrt{1 - A^2} \sin(\pi \Delta z \lambda_e s^2) \right],
$
with reciprocal radius component $s = 2\pi/\sqrt{x^2 + y^2},$ amplitude $A,$ b-factor $b,$ defocus $\Delta z$ and electron wavelength $\lambda_e.$
After applying the point-spread function, we translate the image by $\tau$ and add Gaussian noise with variance $\sigma^2_{\mathrm{noise}} = \sigma^2_{\mathrm{signal}}/\mathrm{SNR},$ where 
 $\sigma^2_{\mathrm{signal}}$ is the variance of the signal and $\mathrm{SNR}$ is the signal-to-noise ratio. The variance of the signal 
$\sigma^2_{\mathrm{signal}}$ is computed by applying a circular mask with a predefined radius on the noiseless image and then calculating the mean squared intensity. The image formation forward model is then
\begin{align}\label{eq:forward model}
   I(x,y | \phi, \rho) = \mathrm{PSF}_{A,b, \Delta z} * &\left(P_z R_q \rho(X) + \boldsymbol{\tau} \right) + \epsilon,  \\ 
   &\qquad  \epsilon \sim \mathcal{N}(0, \sigma^2_{\mathrm{noise}})~, \nonumber
\end{align}
where $*$ denotes convolution.
The imaging parameters utilized for simulating cryo-EM images in CryoSBI are the Gaussian mixture width $\gamma,$ quaternion $q,$ translation $\tau,$ noise level $\sigma_{\mathrm{noise}},$ and PSF parameters $A, b, \Delta z,$ with $\phi = \{\gamma, q, \tau, A, b, \Delta z, \sigma_{noise}\}.$

\section{CryoSBI feature latent network and conditional density estimation}
\label{sbinetworks}

The latent network $S_{\Psi}$ follows a ResNet-18 architecture~\cite{he2016deep} as implemented in ref.~\cite{dingeldein2025amortized}, with modifications for grayscale image input and 256-dimensional feature vector output. For the density estimator $q_{\varphi}$, we implement a Neural Spline Flow (NSF) \cite{durkan_neural_2019} with the same architecture and training as utilized in ref.~\cite{dingeldein2025amortized}, and likewise generating each batch of synthetic images on demand in training.

\section{{ CryoSBI priors}}
\label{priors}

{ All data processing and cryoSBI procedures for the IgG data were carried out as in ref.~\cite{dingeldein2025amortized}. 
The 100 atomic models were generated from an initial structure (PDB id:1HZH) by rotation of a dihedral connecting the fragment antibody (Fab) domain to the rest of the structure (Appendix Figure \ref{fig:igg_atomic_models}), as outlined in ref~\cite{jeon2024cryobench}.
The conformations are indexed by the dihedral angle $\theta_i,
$ with displacement $\theta_{i+1} - \theta_i = 3.6^{\circ}.$}

All processing and SBI for hemagglutinin data were carried out as in ref.~\cite{dingeldein2025amortized}, with
experimental hemagglutinin images obtained from EMPIAR 10532 \cite{tan2020through}, { and whitened using ASPIRE~\cite{aspire} (Appendix Figure~\ref{fig:hemagglutinin}a}). The conformations from hemagglutinin were obtained from a normal mode analysis on atomic structure built from a 3$\text{\AA}$ reconstruction (PDB id: 6wxb), resulting in $20$ conformations indexed by RMSD displacement $\theta_i, i=1,\ldots, 20.$ {(Appendix Figure~\ref{fig:hemagglutinin}b)} 

{ For both datasets},
the conformation prior $p(\theta)$ was taken as a uniform distribution over the possible conformational displacements $\{\theta_i\},$ and the logarithm of the SNR was sampled from a uniform distribution values between $\log 10^{-1}$ and $\log 10^{-3}.$
{ Likewise, for both datasets} the prior on the quaternions $q$ was chosen so that rotations $R_q$ were sampled uniformly in SO(3)~\cite{hanson2005visualizing}. The other imaging parameters were sampled from uniform distributions in each parameter within bounds chosen in ref.~\cite{dingeldein2025amortized}. All nuisance parameters comprising $\phi$ were assumed independent and sampled independently from their respective priors.

\section{Manifold Analysis Framework}
\label{sup_sec:framework}
{
Our manifold analysis as outlined in~\ref{sec:manifold_analysis} follows
a general framework that validates whether $S_\psi$ has learned a representation which exhibits the following desirable properties:
\begin{itemize}
    \item \textbf{Low intrinsic dimensionality}: The encoder should compress images into a low-dimensional manifold with intrinsic dimension $d \ll 256$.
    \item \textbf{Sufficiency for posterior prediction}: The embedding $x = S_\psi(I)$ should retain all relevant information needed by the posterior network $q_\varphi$ to produce accurate estimates of the posterior mean and uncertainty.
    \item \textbf{Invariance to nuisance variation}: The latent representation should be invariant  to uninformative generative factors which do not significantly affect the posterior over $\theta$.
    \item \textbf{Disentanglement of relevant factors}: The embedding should organize generative factors such as conformation or \snr into independent directions in the latent space.
    \item \textbf{Coverage of and fidelity to 
    the experimental data}: The simulator should be able to generate realistic synthetic data that spans the variability observed in experimental images, allowing the neural network ensemble to generalize to new experimental inputs without requiring retraining.
\end{itemize}

Our analysis focuses on the statistical and geometric properties of the embedding produced by $S_\psi$, offering a complementary validation procedure to works such as \cite{talts2020validatingbayesianinferencealgorithms} which focus on the calibration of the estimated posterior distribution. Our framework could be viewed as a model misspecification detection protocol, and in future work could be synergized with other analysis of latent embeddings in neural posterior estimation~\cite{schmitt2024detecting, huang2023learning}.




An overview of our pipeline can be found in 
Appendix Figure \ref{fig:framework}. 
If a step in our framework is applicable to both synthetic and experimental data, we suppress $s$ or $e$ from the subscript(e.g. we write $\M$ instead of $\M_e$). 
}
\section{Data Pre-processing}
\label{sup_sec:pre-processing}
Our data pre-processing consists of three steps: subsampling, outlier removal and resampling the data. Here we provide information on each step, with general guidelines and implementation details for the datasets in this work.

\subsection{Data Sub-sampling}\label{sup_subsec:data-subsample}

{ Most stages of our framework require computations that scale as $\mathcal{O}(N^2)$, primarily due to pairwise distance or neighbor graph construction which can be prohibitive for large $N$. Thus, if needed, we can randomly sub-sample the data down to a more manageable size. However, the sub-sample must be large enough to capture the underlying geometric structure, maintain statistical robustness, and ensure that we can uniformly subsample the data later in the pipeline, but small enough to remain tractable given available computational resources. In practice, we find that selecting approximately three times the number of points that will be used in the final manifold learning step is a reasonable balance. The manifold embedding itself should be performed with as many points as the practitioner’s computational budget allows, since more points typically yield a better approximation of the underlying geometry. In our experiments, we use $20,000$ points for computing the Diffusion Maps embedding, so we subsample all datasets down to $60,000$ points. }


\subsection{Outlier Removal} \label{sup_subsec:outliers}

{ Outlier detection serves a critical function in our validation framework as it removes pathological or noisy points that can distort downstream geometric analysis(e.g. spectral embedding, intrinsic dimension estimation). Thus, for each dataset $\X$, we train an outlier detector $\hat{y} : \R[256] \to \{0, 1\}$, which classifies input neural embeddings as either in the distribution $p$ or out of the distribution $p$. While various methods exist - including One-Class SVMs~\cite{schölkopf2001OCSVM}, Isolation Forests~\cite{liu2008IsolationForrest}, or KDE-based detectors~\cite{lang2022KDEOutlier} - we favor \textit{Minimum Volume Sets (MVS)}~\cite{scott2006learning, saligrama2009MVS} for their conceptual simplicity, strong theoretical guarantees, and alignment with the local neighborhood statistics used throughout our pipeline. 

An $\alpha$-MVS is the smallest measurable set $U_\alpha$ containing at least a fraction $\alpha$ of the probability mass of $p$. Points lying outside of $U_\alpha$ are flagged as outliers, with the magnitude of $1 - \alpha$ determining the stringency of the anomaly filtering. As shown in~\cite{saligrama2009MVS}, one can use an empirical sample $\X \sim p$ to construct a simple membership test that, in the limit, consistently approximates $U_\alpha$. Specifically, we approximate the local density at each point using either the inverse $k$-nearest neighbor distance, $1/\dist(x)$, or the neighbor count $n_r(x)$ within radius $r$ of $x$, both of which are proportional to the true local density. Then, given a fixed number of neighbors $k$, the classifier(i.e. the $U_\alpha$ membership test) is defined as:
\[
\hat{y}_k(x') := 
\begin{cases}
1, & \text{if } 1 - \alpha \leq s_k(x') \\
0, & \text{otherwise}
\end{cases}
\]
for \( s_k(x') := \frac{1}{N}\sum_{x \in \X} \mathbb{I}[\dist(x') \leq \dist(x)] \)(or an analogous formula if using $n_r$ with fixed radius $r$ and with reversed inequality signs). Intuitively, this means that a point is deemed in-distribution if its estimated local density exceeds that of at least a fraction $1-\alpha$ of the training points. For robustness, we can instead use the classifier:
\[
\hat{y}(x') := 
\begin{cases}
1, & \text{if } 1 - \alpha \leq \frac{1}{|K|}\sum_{k \in K}s_k(x') \\
0, & \text{otherwise}
\end{cases}
\]
which averages the score $s_k(x')$ over multiple values of $k \in K$(and similarly for $n_r$ by averaging over multiple radii $r \in R$). We obtain the `clean' dataset $\X^{clean} := \{x \in \X \; | \; \hat{y}(x) = 1\}$ of size $N^{clean} = \mathrm{round}(N \cdot \alpha)$. 

In our experiments, we apply the robust MVS outlier filter defined above to all three of our datasets. For the \ig dataset, we use $\alpha = 0.2$. For the \hem datasets, we use $\alpha = 0.3$. For all datasets we average the scores for $K \in \{10, 20, \ldots, 120\}$. In Appendix Figure~\ref{fig:alpha_selection} we provide practical guidance on selecting $\alpha$.

}

\subsection{Uniform Resampling}\label{app:uniform_resampling}

{ Many manifold learning methods, including our intrinsic dimensionality estimators(Section~\ref{subsec:dim_estim}) and Diffusion Maps(Section~\ref{subsec:embed}), either explicitly assume or empirically perform better when the distribution over the underlying manifold is uniform. For example, a non-uniform distribution $p$ over $\M$ introduces bias in the estimation of the Laplace–Beltrami operator $\Delta_\M$ via Laplacian Eigenmaps~\cite{belkin2001LapEig}. While the normalization scheme used by Diffusion Maps~\cite{coifman2005geometric} theoretically corrects this bias, the result is asymptotic and relies on mild density variations, which rarely hold in practice.

To address this, we perform inverse density-weighted resampling to produce a subsample that approximates uniform coverage of the manifold. Empirically, we observe a significant improvement in performance after implementing this technique. Additionally, a uniform distribution reduces the need for adaptive bandwidths in local neighborhood computations and enables reliable use of a single global kernel width $\epsilon$. 

For fixed $k \in \mathbb{N}$, we estimate the local density at each  $x \in \X^{clean}$ using the nonparametric estimator in \cite{quesenberry1966nonparametric}:
\[
\hat{p}_k(x) = 
\frac{k}{N^{clean} \cdot V_{d}(\dist(x))}
\]
which is consistent under mild conditions~\cite{boente1988ConsistencyNonParametric}. Here $\dist(x)$ is the distance to the $k$-th nearest neighbor, $d$ is the intrinsic dimension of $\M$, and $V_{d}(r)$ is the volume of a $d$-dimensional ball of radius $r$. We obtain local estimates of $d$ by the \textit{doubling dimension} method~\cite{assouad1983doubling} described in Section \ref{subsec:dim_estim}. For robustness, we compute density estimates for a range of $k \in K$ values and average the results obtaining the density estimate:
\[
\hat{p}(x) = \frac{1}{|K|}\sum_{k \in K} \hat{p}_k(x)
\]
In our experiments we use $K \in \{10, 20, \ldots, 120\}$. Practitioners are free to choose any local density estimator they prefer; our choice is motivated by conceptual simplicity, effectiveness, and by the fact that we end up using the same local statistics through out our pre-processing routine.  

After estimating local densities, we assign sampling weights proportional to the inverse density. Normalizing these weights yields a probability distribution over $\X^{clean}$ from which we draw, without replacement, a new subset $\X^{unif}$ of size $N^{unif}$. This resampled dataset approximates a uniform sample on the manifold, effectively disentangling geometry from the sampling distribution. For all our experiments we use $N^{unif} = 20,000$. It is very important that $N^{clean}$ is large enough relative to $N^{unif}$ in order to allow the sub-sampled distribution to shift towards a uniform one. As an extreme example, if $N^{clean} = N^{unif}$, then the sub-sample will have exactly the same distribution $p$ as $\X^{clean}$ as we are sampling without replacement.
}

\section{Additional Details for Intrinsic Dimension Estimation}\label{sup_sec:dim_estim}

{
This section provides more detail about the four intrinsic dimensionality algorithms employed in our study. We will focus more on practical implementation specifics, including hyperparameter selection and diagnostic procedures used in our experiments, and recommend the review by~\cite{camastra2016IntrinsicDim} for broader theoretical context.

\textbf{Correlation Dimension}\citep{grassberger1983slope}:
This method relies on the observation that the number of neighbors within a radius $r$ scales proportionally to $r^d$, where $d$ is the intrinsic dimension. As such, $\log n_r(x) = d \cdot \log r + \mathrm{const}$. So, if we compute $\overline{n_r} = \frac{1}{N^{unif}} \sum_{x \in \X^{unif}} n_r(x)$, the average number of neighbors over the whole dataset for $r$ in a range of radii $R$, we can globally estimate $\hat{d}$ as the slope of the line fitted to $\log \overline{n_r}$ versus $\log r$.

In our experiments, we use $R = \{8.0, 8.5, 9.0, \dots, 18.0\}$. The log-log plot should appear linear as in (Figure~\ref{fig:d_estim}a). Downward curvature at the ends indicates $R$ is too wide, and extreme values should be removed from the range $R$. Otherwise, the predicted $\hat{d}$ will likely be biased too low.

\textbf{Eigengap Method}\citep{chen2013eigengap}: This approach exploits the idea that the eigenvalues $\{\lambda_k\}_{k=1}^{256}$ of the local covariance matrix are computed from data sampled from a $d$-dimensional manifold embedded in $\R[256],$ and thus should exhibit a significant gap between $\lambda_d$ and $\lambda_{d+1}$. The weighted local covariance at a point $x_i \in \X^{unif}$ is computed as:
\[
C_{x_i}= \sum_{x_j \in \mathcal{N}_r(x_i)}\frac{w_{ij}}{w_i}(x_j - x_i)^\top(x_j - x_i) 
\]
where $ \mathcal{N}_r(x_i)$ denotes neighbors within a radius $r$ of $x_i$, $w_{ij} = \exp\left(-\frac{\|x_i - x_j\|_2^2}{\epsilon^2}\right)$, and $w_i = \sum_{x_j \in \mathcal{N}_r(x_i)} w_{ij}$. We perform the truncated SVD of $C_{x_i}$ for a conservatively high upper bound $D > d$ in order to obtain the eigenvalues. In our experiments we use $D = 20$. Parameters $\epsilon$ and $r$ are selected via the distortion-minimization procedure of \cite{perraultjoncas2014consistency} which in our experiments we run over values $r \in \{8.0, 9.0, \dots, 22.0\}$ with $\epsilon = r / 3$, a typical default bandwidth.
For computational efficiency, we compute the local dimension at a subsampled set of $5000$ data points for each dataset.
The local dimension is estimated using either the maximum gap or the softmax gap:
\[ 
\hat{d}(x_i) = \arg\max_{k} g_k \quad \text{ or } \quad \hat{d}(x_i) = \frac{\sum_{k=1}^{D-1} k \cdot e^{g_k}}{\sum_{k=1}^{D-1} e^{g_k}}
\]
for $g_k = \lambda_k - \lambda_{k+1}$. We prefer the softmax-weighted version as it smooths out the prediction, making it more robust to noise. A global prediction $\hat{d}$ is obtained by averaging local estimates $\hat{d} = \frac{1}{N^{unif}}\sum_{x \in \X^{unif}}\hat{d}(x)$.

\textbf{Doubling Dimension}\citep{assouad1983doubling}:
Similar to correlation dimension, this method examines how point counts scale with radius, but focuses specifically on the correlation when doubling the radius. The local estimate for a fixed radius $r \in \R^+$ is:
\[
\hat{d}_r(x) = \log_2 \left( \frac{n_{2r}(x)}{n_r(x)} \right)
\]
which we can average over a range of radii $R$ as $\hat{d}(x) = \frac{1}{|R|} \sum_{r \in R} \hat{d}_r(x)$ in order to improve robustness. A global prediction $\hat{d}$ is obtained by averaging local estimates $\hat{d} = \frac{1}{N^{unif}}\sum_{x \in \X^{unif}}\hat{d}(x)$.

In our experiments we use $R = \{8.5, 9.0, \dots, 11.0\}$. An important practical aspect is that the estimates should be stable across the chosen range $R$ as seen in
Appendix Figure~\ref{fig:dd_all}). If there is significant variation, then the range $R$ should be tightened.

\textbf{Levina–Bickel MLE}\citep{levina2004MLEdim}:
This estimator uses the scaling of distances to nearest neighbors. For a fixed $k \in \mathbb{N}$ we have the local estimates:
\[
\hat{d}_k(x) = \left[ \frac{1}{k-1} \sum_{j=1}^{k-1} \log \frac{\dist[k](x)}{\dist[j](x)} \right]^{-1}
\]
where $\dist[j](x)$ is the distance to the $j$-th nearest neighbor. We replace the $k-1$ denominator with $k-2$ in order to obtain an unbiased estimator as shown in the original paper. Similar to previous methods, we can obtain a robust estimator $\hat{d}(x)$ by averaging $\hat{d}_k(x)$ over multiple values of $k$ in a range $K$ and a global estimate $\hat{d}$ by averaging these values in turn.

We use $K = \{10, 20, \dots, 120\}$. As with the doubling dimension,  in practice the estimates should be stable across $K$ as in 
Appendix Figure~\ref{fig:lb_all}, and if one sees significant variation, the range of $K$ should be smaller.
}

\section{Additional Details on KDE} 
\label{sup_sec:cover}
{
For kernel density estimation (KDE), 
we sample disjoint training, validation, and testing data sets from $\X_e^{clean}$ and $\X_s^{clean}$ as obtained by
outlier removal in Appendix~\ref{sup_subsec:data-subsample}.
Note that we are not using the uniformly resampled $\X_e^{unif}, \X_s^{unif}$ for this purpose as they are not samples from the data distributions $p_e$ and $p_s$. The training and validation sets will be used to fit the KDE models and to select appropriate kernel bandwidths, while the test subset will be used in the Monte-Carlo estimation of the divergence.
In order to train the KDEs and evaluate the Monte Carlo estimated KL divergences, we sample the following sample subsets from $\X_s^{clean}$ and $\X_e^{clean}$:
\begin{itemize}
  \item \textbf{Training set:} used to fit each KDE. Size 17,000 in our experiments.
  \item \textbf{Validation set:} used for bandwidth selection. Size 3,000 points in our experiments. We select the kernel bandwidths $h_s$ and $h_e$ that maximize the log-likelihood of the validation set. We use a single validation run on the whole validation set
  to minimize. We initially search over a wide exponential grid to find a coarse range, then tighten the search interval. The final bandwidths used are: $h_s = 0.48$ and $h_e = 0.31$.
  \item \textbf{Test set:} an independent subset for Monte Carlo estimation of the divergences. Size 3,000 points in our experiments. 
\end{itemize}
}
\section{Manifold Learning Details}\label{sup_sec:embed}

{
Our manifold learning step first requires construction of an affinity matrix.
We follow the method of~\cite{perraultjoncas2014consistency} to estimate the kernel bandwidth $\epsilon$ and cutoff radius $r$ that best preserve the geometry of the manifold when constructing the Gaussian affinity matrix. We perform the search over radii $r \in \{8.0, 9.0, \dots 22.0\}$ and set $\epsilon = r/3,$  a typical default value. For the \hem dataset we find $r_e = 14.0$ for the experimental data and $r_s = 13.0$ for the simulated data. For the \ig dataset, we obtain $r_s = 15.0$. After the initial estimation, we remove all latent data vectors whose degrees in the affinity matrix fall in the bottom 5\%. The outlier removal step improves the stability of the eigen-decomposition in the Diffusion Maps algorithm, and is generally a useful step for whatever embedding algorithm is used.
Alternatively, the user could check a histogram of the degrees in the affinity matrix and select a threshold that removes a long tail of nodes with very small degrees. 

Given the filtered affinity matrix, we compute the Diffusion Maps embedding~\citep{coifman2005geometric} by performing an eigen-decomposition of the geometric Laplacian $\LL$, which, in the limit, estimates the Laplace-Beltrami operator $\Delta_\M$. In our experiments we retain the top $m = 20$ non-trivial eigenvectors to form the initial embedding matrix $\Phi$. The value of $m$ should be comfortably larger than what we expect the real $d$ to be, but not excessively so. The eigenvalue spectra, shown in Appendix Figures~\ref{fig:hem_eigvals} and~\ref{fig:igg_eigvals}, confirm that only the first eigenvalue is zero and that the spectra grow slowly, indicating that the manifolds are smooth, connected, and free of isolated clusters.

To ensure that the final coordinates used for visualization and interpretation are truly informative and non-redundant, we apply Independent Eigencoordinate Selection (IES)~\citep{chen2019selecting} on the initial spectral embedding. The key problem IES addresses is that Diffusion Maps (or any spectral embedding) may yield embedding coordinate functions that are dependent. For example, a pair like $(x, \sin(x))$ varies in two directions but really contains only one independent degree of freedom. Visualizing both would be misleading or wasteful if the goal is to reveal distinct generative factors. To solve the problem, IES searches for a subset $S$ of the initial embedding coordinates $\Phi$ that maintains full local rank $d$ everywhere on the manifold, thus avoiding singularities. In our usage, we set the subset size $|S|$ equal to our estimated intrinsic dimension $\hat{d}$, but choose the target rank $d$ to match the number of known, highly relevant generative factors we expect the manifold to express($\theta$ and \snr for our datasets). 
For computational efficiency we subsample all input datasets to a set of $500$ points for computing IES.
Our final selected subsets of embedding coordinates are $S_s = \{0, 2, 3\}$ for the experimental \hem data and $S_e = \{0, 1, 2, 3, 4\}$ for the simulated version, while for the \ig dataset we obtain $S_s = \{0, 1, 2, 5, 6, 14\}$. We then display the first three coordinates which for DM embeddings are the lowest frequency ones. 

As an optional refinement, we apply Riemannian Relaxation \cite{mcqueen2016relaxation} to push the embeddings closer to being isometric to their respective neural representations. To do this, Riemannian Relaxation starts from the initial embeddings $\Phi^{indep}$ and iteratively modifies them via gradient descent with respect to a loss function which penalizes local distortions. For this work, we only applied Riemannian Relaxation to the \hem datasets. In Appendix Figure \ref{fig:relaxation}, we display "relaxed" versus "unrelaxed" versions of $\Phi_e$ and $\Phi_s$ for the \hem data. For the \hem datasets, we use $d=3$ and $\epsilon_{orth} = 0.5$, and run Riemannian Relaxation until convergence. 
}

\section{Estimating the parameters of the experimental data by interpolation}\label{sup_sec:interpolation}

In this section, we explain how we infer the generative parameters $\tilde{\theta}$ and $\tilde{\phi}$ for the experimental data and how we embed a new sample from $\mathcal{X}_e$ into the embedding space $\Phi_e$ as in Appendix Figure \ref{fig:exp_density}. This is done via Nadaraya-Watson Kernel Regression \cite{murphy2022probabilistic} in the neural embedding space. More specifically, for every $\tilde{x}_i \in \mathcal{X}_e$, we estimate the conformation $\tilde{\theta}_i = \frac{\sum_{x_j \in \mathcal{X}_s} K(\tilde{x}_i, x_j)\theta_j}{\sum_{x_j \in \mathcal{X}_s} K(\tilde{x}_i, x_j)}$. Similarly, we obtain estimated nuisance parameters $\tilde{\phi}_i$. To embed a new point $\hat{x}_i \in \mathcal{X}_e$ in the embedding space $\Phi_e$, we compute the $c$-th coordinate of $\Phi_e(\hat{x}_i)$ as $\Phi^c_e(\hat{x}_i) = \frac{\sum_{\tilde{x}_j \in \mathcal{X}_e} K(\hat{x}_i, \tilde{x}_j)\Phi^c_e(\tilde{x}_j)}{\sum_{\tilde{x}_j \in \mathcal{X}_e} K(\hat{x}_i, \tilde{x}_j)}$.

\section{Gradient estimation}\label{sup_sec:grad_estim}

{
Since the gradients $\nabla f_k$ are not analytically known, we estimate them using a local finite-difference procedure. For each point $x \in \X^{unif}$, we compute a weighted local neighborhood based on the same kernel matrix used for Diffusion Maps. We perform local weighted PCA to obtain a local basis $U(x) \in \R^{256 \times d'}$, where we use $d' = 10$ principal directions. We then construct local differences of coordinates and function values for neighbors $x'$ of $x$ respectively as:
\begin{align*}
    &\Delta_x(x') = w(x')(x' - x) \in \R[256]  \\
    &\Delta_{f_k}(x') = w(x')(f_k(x') - f_k(x)) \in \R
\end{align*}
where $w(x')$ is the kernel weight for neighbor $x'$. Next, we project the coordinate differences unto the span of $U(x)$ as $\widetilde{\Delta}_x(x') = [U(x)]^\top\Delta_x(x') 
\in \R[d']$. From all neighbors of $x'$ of $x$, we gather all $\widetilde{\Delta}_x(x')$ into a design matrix and all $\Delta_{f_k}(x')$ into the target vector and solve the resulting weighted least-squares problem for $g \in \R[d']$. This is an estimate of the gradient represented in $U(x)$ coordinates which we transform into an estimate in the original space by
\(\nabla f_k(x) = U(x) g\).

The resulting gradients are then used as input to TSLasso. We note that the projection step is optional, but we found it useful both in terms of robustness and computational efficiency as it reduces the dimensionality of the least squares problem from 256 features to $d' \ll 256$.  

}

\section{TSLasso Details}\label{sup_sec:tslasso}

{
We use TSLasso~\citep{koelle2024consistency} to validate whether the embeddings learned by $S_\psi$ are predictive for inferring the conformation posterior while remaining robust to nuisance variation. TSLasso performs feature selection on manifolds: given a dictionary $\mathcal{F} = \{f_k: \M \to \R\}_{k=1}^p$ of candidate functions—here, the simulated generative parameters such as conformation $\theta$ and nuisance factors $\phi$—it searches for a subset of $\mathcal{F}$ whose gradients $\{\nabla f_k\}^p_{k=1}$ best span the tangent bundle of the manifold $\M$.

First, TSLasso estimates local tangent spaces from the data using local PCA. Then, it projects the gradients $\nabla f_k$ onto these tangent spaces, and finally reconstructs local bases as sparse linear combinations of the projected gradients. The magnitudes of the linear coefficients $B_k$ are regularized with a Group Lasso penalty to encourage sparsity across all samples. We sweep the regularization parameter $\lambda$ to find a value for which exactly $d$ dictionary elements have corresponding non-zero coefficients $B_k$. The average magnitudes of $B_k$ as $\lambda$ varies indicate how strongly each function contributes to spanning the local geometry.

In our experiments, each $f_k$ corresponds to one of the known generative parameters (the conformation $\theta$ or one of the nuisance factors in $\phi$), giving $|\mathcal{F}| = 10$ for all datasets. For the \hem dataset we use $|S| = 4$; for \ig we use $|S| = 5$, values which are slightly lower than the estimated $\hat{d}$'s for the experimental \hem dataset and the synthetic \ig datasets, but within a reasonable range. The true generative factors are unknown for experimental data, so we infer them by kernel interpolation from nearby simulated data, yielding $\tilde{\theta}$ and $\tilde{\phi}$ (Appendix~\ref{sup_sec:interpolation}).

We run TSLasso 20 times using random subsets of 500 points drawn from the $\X^{unif}$ sample. To test robustness to noise, each run samples points whose $\mathrm{SNR}$ (inferred for experimental data) values are within the top $q$-th percentile of all points, with $q \in \{0, 5, \dots, 95\}$. The results are summarized in Appendix Figure~\ref{fig:tslasso}. We find that TSLasso almost always selects conformation $\theta$ (or $\tilde{\theta}$ for experimental data) and $\mathrm{SNR}$(\snrt for experimental data) as the dominant factors parametrizing the embedded manifold.
}

\section{Runtime Considerations}\label{sup_sec:runtime}

All operations in our framework that involve pairs of neighbors( neighbor counts, distances to the $k$-th nearest neighbor, KDE distances, etc.), as well as the spectral decomposition of the Laplacian matrix $\mathbf{L}$ required to compute the embeddings, scale as $\mathcal{O}(N^2)$ for $N$ the number of points used for that computation. Our experiments start at $N = 60000$ and after uniform sub-sampling reduces to $N = 20000$. We implement these operations using sparse linear algebra NumPy routines, which greatly reduces both time and space complexity.
The experiments involving the computation of neighbor counts, neighbor distances and KDE distances as described in Appendix~\ref{sup_sec:pre-processing}, were all performed in under 10 minutes on a personal laptop.

The remaining operations of gradient estimation, Eigengap dimensionality estimation, IES, and TSLasso, require point-wise estimations of tangent spaces which are obtained by spectral decompositions of local covariance matrices. Thus, these operations scale as $\mathcal{O}(N^3)$. However, we need not compute these local decompositions on the whole data set, but only on a representative sub-sample. For the Eigengap dimension estimate and IES, we used $5000$ points, while for IES and TSLasso we used $500$ points. Using personal laptops, these operations amounted to under 20 minutes. 

Finally, Riemannian Relaxation involves a heavy iterative optimization procedure and is by some margin the most expensive operation in our framework taking up to 3 hours. Fortunately, this procedure is an optional step for the pipeline, and future work can investigate the use of early stopping to reduce the computational load.

\renewcommand{\figurename}{Appendix Figure}
\setcounter{figure}{0}
\begin{figure*}[ht]
    \centering
    \includegraphics[width=\textwidth]{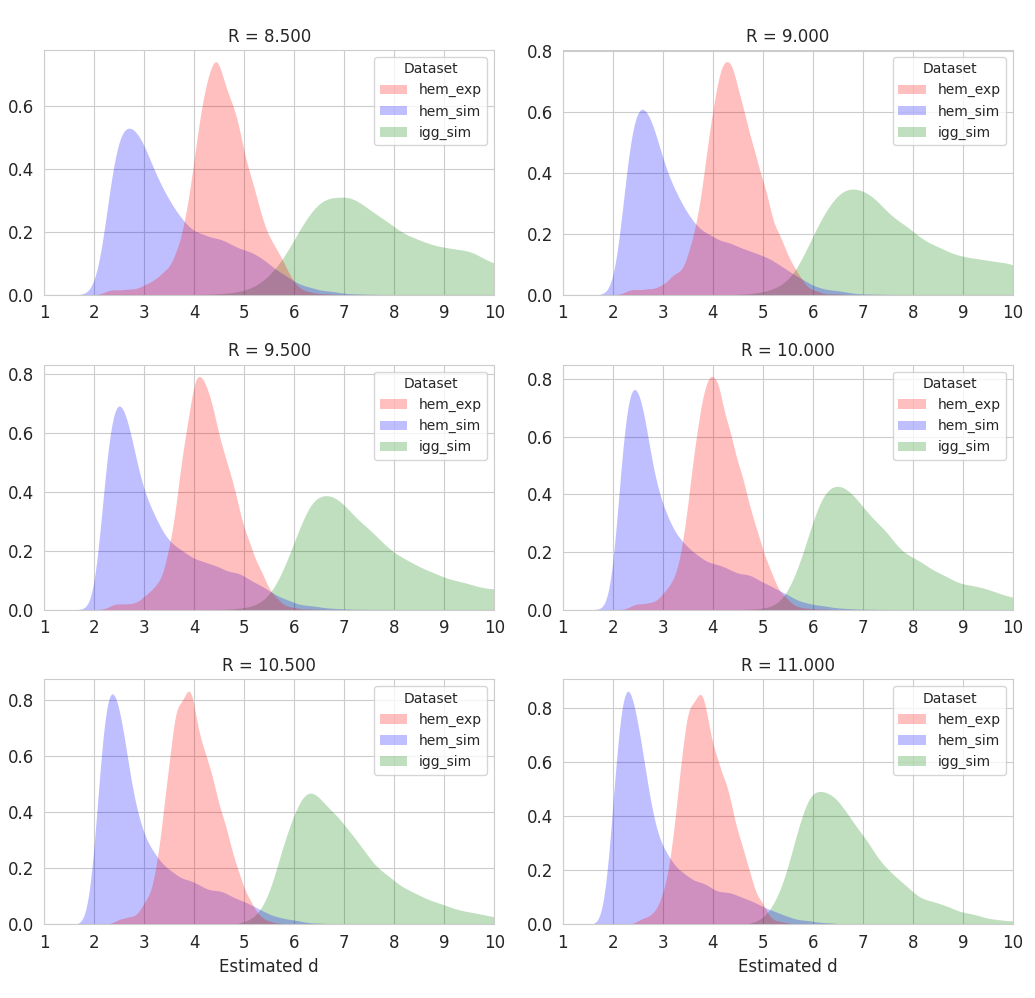}
    \caption{
    { Point-wise estimates of the intrinsic dimension $d$ using the Doubling Dimension method for different radii $r$ for the experimental (red) and synthetic (blue) \hem data manifolds, and for the synthetic \ig data manifold (green).}
    }
    \label{fig:dd_all}
\end{figure*}

\begin{figure*}[ht]
    \centering
    \includegraphics[width=\textwidth]{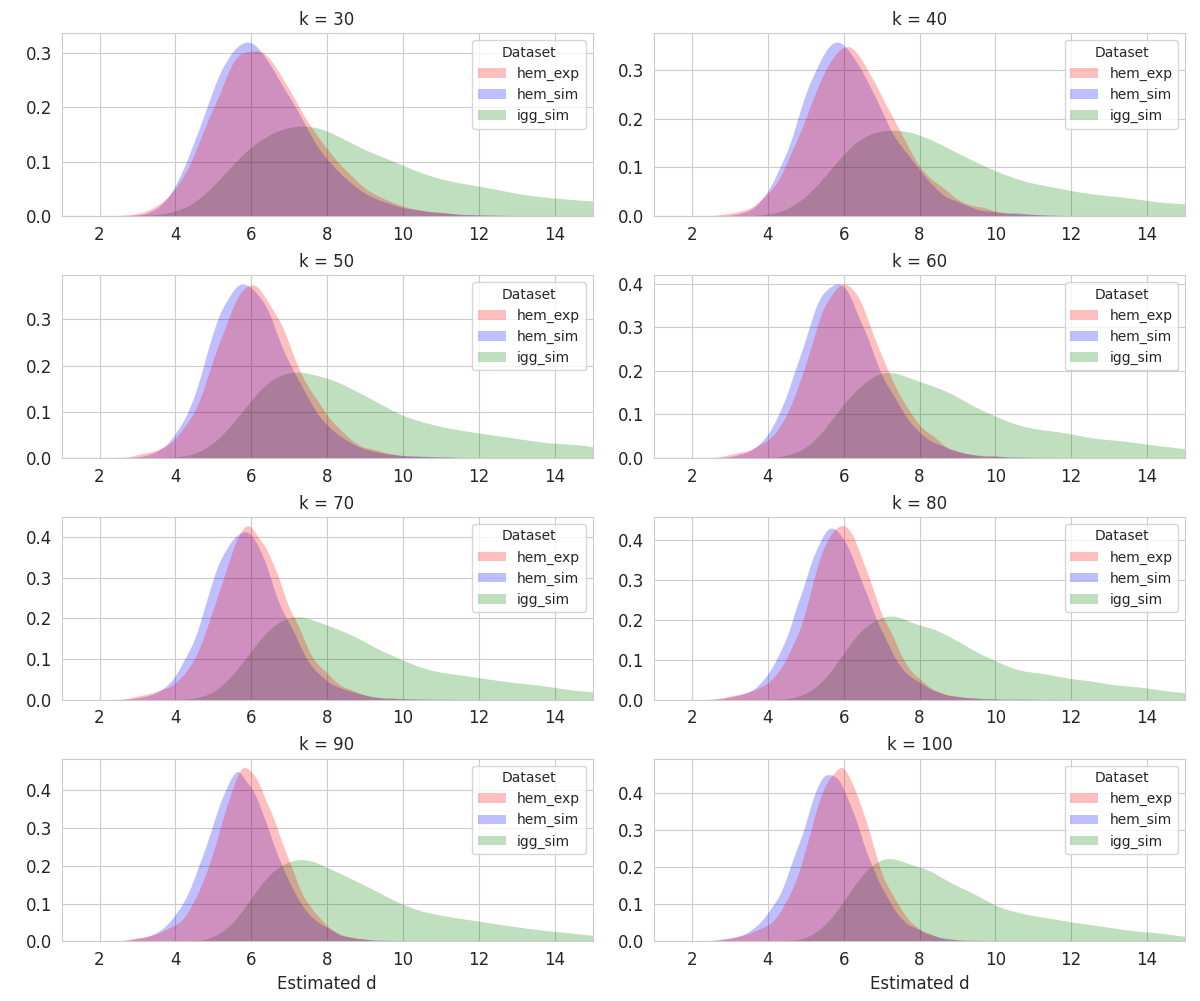}
    \caption{
    { Point-wise estimates of the intrinsic dimension $d$ using the Levina-Bickel MLE for different $k$-th nearest neighbor distances for the experimental (red) and synthetic (blue) \hem data manifolds, and for the synthetic \ig data manifold (green).}
    }
    \label{fig:lb_all}
\end{figure*}

\begin{figure*}[ht]
    \centering

    \subfloat[$\Phi_s$ colored by $\theta$.]{\includegraphics[width=0.35\textwidth]{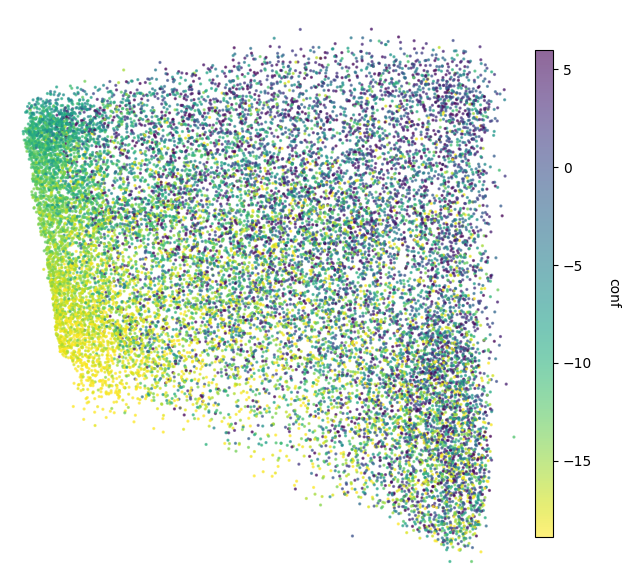}}
    \hspace{3cm}
    \subfloat[$\Phi_s$ colored by $\hat{\theta}$.]{\includegraphics[width=0.35\textwidth]{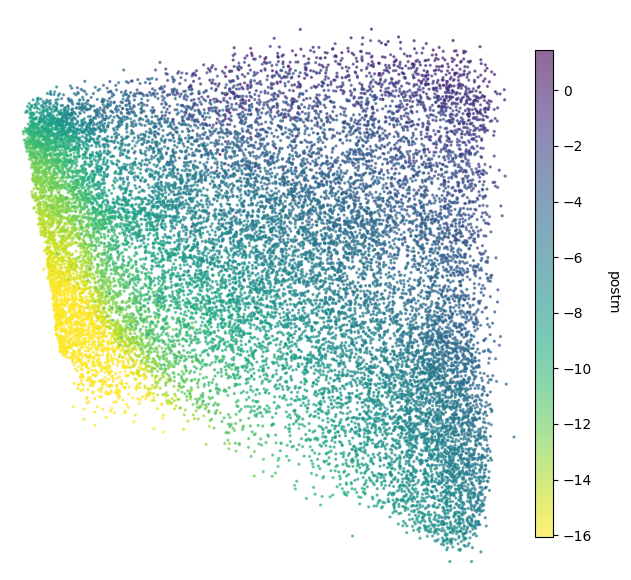}}

    \vspace{0.1cm}

    \subfloat[$\Phi_s$ colored by $\mathrm{SNR}$.]{\includegraphics[width=0.35\textwidth]{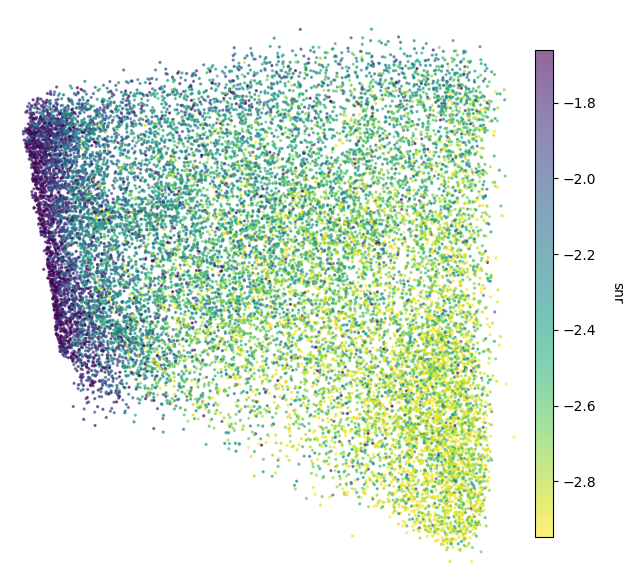}}
    \hspace{3cm}
    \subfloat[$\Phi_s$ colored by $\sigma$.]{\includegraphics[width=0.35\textwidth]{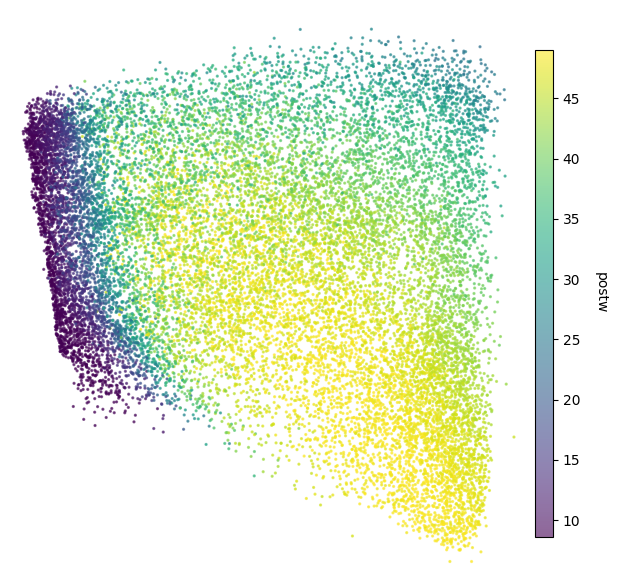} }

    \vspace{0.1cm}

    \subfloat[$\Phi_s$ colored by $d_s$ estimated by the Eigengap method.]{\includegraphics[width=0.35\textwidth]{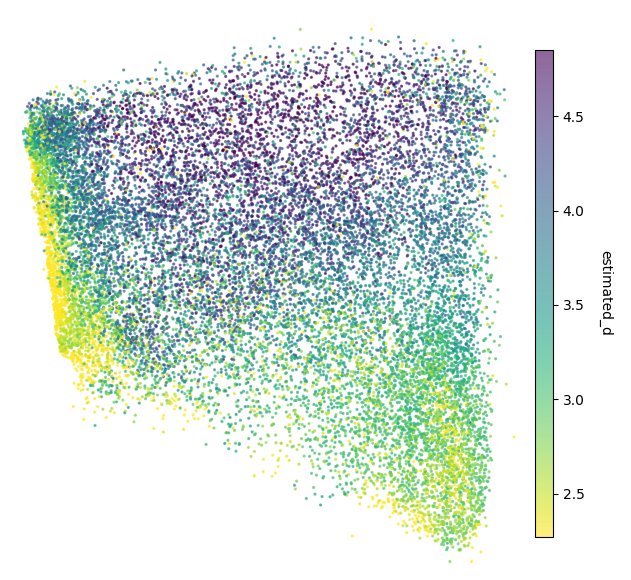}}
    \caption{
        \raggedright
        Diffusion Maps embeddings of the simulated \hem data $\Phi_s$ in $d=3$ dimensions. The plots are rotated to best display the embedding. The three coordinates we display are selected by IES.
        In \textbf{(a)} and \textbf{(b)}, for data points with high SNR (on the left-hand side of the point cloud in both figures), the conformation and posterior mean agree and vary smoothly across the y-axis. 
        In \textbf{(c)} and \textbf{(d)}, the SNR and posterior width agree over the embedded points and vary smoothly across the x-axis, plotted with opposing color scales to emphasize regions where SNR and posterior width agree.
        In \textbf{(e)}, the intrinsic dimension is highest in regions with medium SNR. For data with high SNR (the left most points), the intrinsic dimension $d_s$ drops due to the lack of noise; for the noisiest data (lower right of embedding), $d_s$ drops again, as noisy images become more similar to each other.
        }
    \label{fig:sim_embeddings}
\end{figure*}

\begin{figure*}[ht]
    \centering
    
    \subfloat[$\Phi_s$ colored by $\theta$.]{%
        \includegraphics[width=0.45\textwidth]{Figures/igg_conf.png}
    }
    \hfill
    \subfloat[$\Phi_s$ colored by $\hat{\theta}$.]{%
        \includegraphics[width=0.45\textwidth]{Figures/igg_postm.png}
    }

    \subfloat[$\Phi_s$ colored by SNR.]{%
        \includegraphics[width=0.45\textwidth]{Figures/igg_snr.png}
    }
    \hfill
    \subfloat[$\Phi_s$ colored by $\sigma$.]{%
        \includegraphics[width=0.45\textwidth]{Figures/igg_postw.png}
    }

    \subfloat[$\Phi_s$ colored by $d_s$ estimated by the Doubling Dimension method.]{%
        \includegraphics[width=0.45\textwidth]{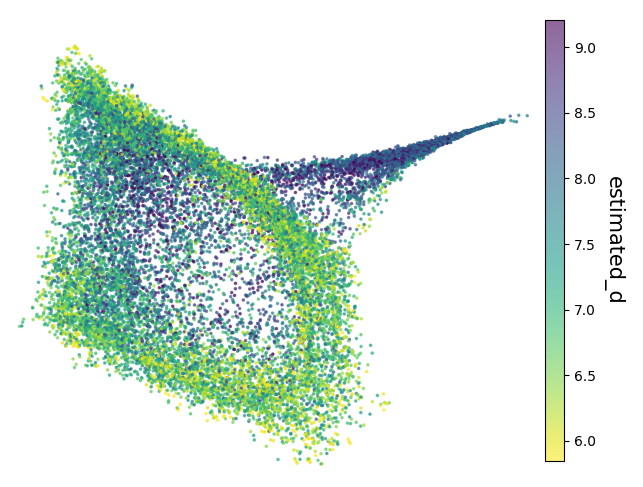}
    }

    \caption{
    \raggedright
    { Diffusion Maps embeddings $\Phi_s$ in $d=3$ dimensions for the IgG dataset. The plots are rotated to best display the embedding. The three coordinates we display are selected by IES. 
    In \textbf{(a)} and \textbf{(b)}, for data points with high SNR (on the ``ring'' of the point cloud in both figures), the conformation and posterior mean agree and vary smoothly, except for a boundary artifact due to periodicity of $\theta$. 
        In \textbf{(c)} and \textbf{(d)}, the SNR and posterior width agree over the embedded points and vary smoothly from the ``tail'' of the point cloud (low SNR) to the ``ring'' (high SNR)  , plotted with opposing color scales to emphasize regions where SNR and posterior width agree.
        In \textbf{(e)}, the intrinsic dimension is lowest in regions with high SNR, and higher towards the ``tail'' region of low SNR.}
    }\label{fig:igg_embedding}
\end{figure*}

\begin{figure*}[ht]
    \centering

    \subfloat[$\Phi_e$ colored by $\theta$.]{\includegraphics[width=0.35\textwidth]{Figures/exp_conf.png}}
    \hspace{3cm}
    \subfloat[$\Phi_e$ colored by $\hat{\theta}$.]{\includegraphics[width=0.35\textwidth]{Figures/exp_postm.png}}

    \vspace{0.1cm}

    \subfloat[$\Phi_e$ colored by $\mathrm{SNR}$.]{\includegraphics[width=0.35\textwidth]{Figures/exp_snr.png}}
    \hspace{3cm}
    \subfloat[$\Phi_e$ colored by $\sigma$.]{\includegraphics[width=0.35\textwidth]{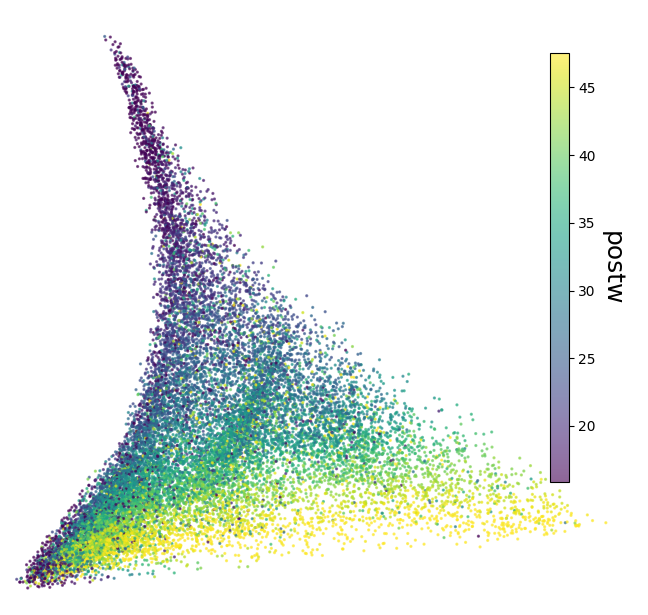} }

    \vspace{0.1cm}

    \subfloat[$\Phi_e$ colored by $d_s$ estimated by the Eigengap method.]{\includegraphics[width=0.35\textwidth]{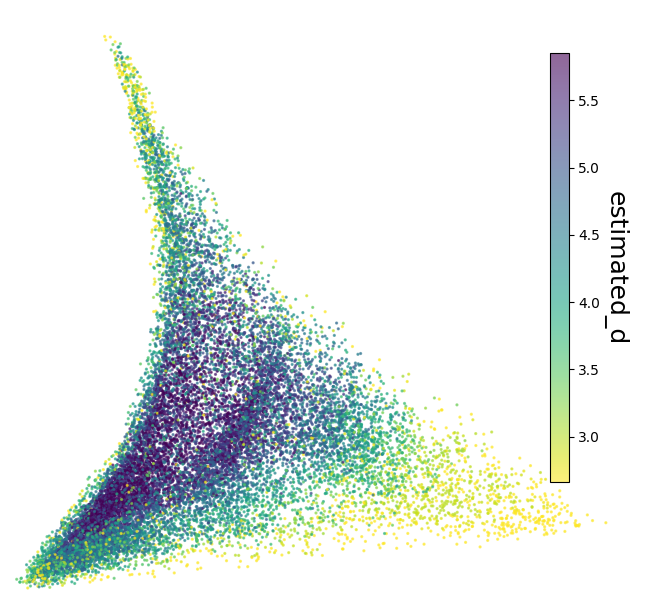}}

    \caption{
    \raggedright
    { Diffusion Maps embeddings $\Phi_e$ in $d=3$ dimensions for the experimental \hem dataset. The plots are rotated to best display the embedding. The three coordinates we display are selected by IES. In   
        \textbf{(a)} and \textbf{(b)}, $\Phi_e$ is colored by the predicted conformation from manifold interpolation $\tilde{\theta}$  and the conformation estimated posterior mean $\hat{\theta}$ respectively. In plots \textbf{(c)} and \textbf{(d)} 
         $\Phi_e$ is colored by the interpolated SNR and posterior width $\sigma$ respectively, with opposing color scales to emphasize that high SNR regions are often similar to low posterior width regions and likewise for low SNR and high posterior widths.
In \textbf{(e)} $\Phi_e$ colored by local $d_e$. The highest intrinsic dimension is in regions with medium SNR, while high SNR regions have $d_e \in [3,4]$. 
}}
\label{fig:exp_embeddings}
\end{figure*}

\begin{figure*}
    \includegraphics[width=\textwidth]{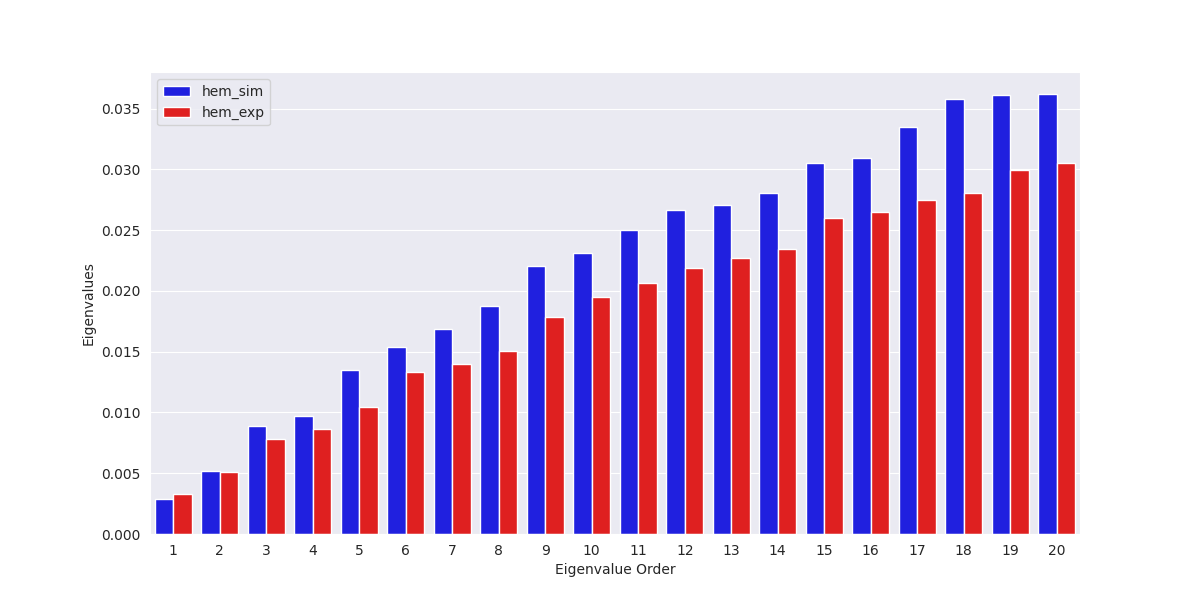}
    \caption{
    { The spectrums of the eigen-decompositions of the Diffusion Maps Laplacian matrices $\LL$ computed for the \hem experimental (red) and simulated (blue) datasets. The smoothness of the spectrum and having only one 0 eigenvalue(not displayed) indicates that both $\mathcal{M}_s$ and $\mathcal{M}_e$ are smooth connected manifolds.}
    }
    \label{fig:hem_eigvals}
\end{figure*}

\begin{figure*}
    \includegraphics[width=\textwidth]{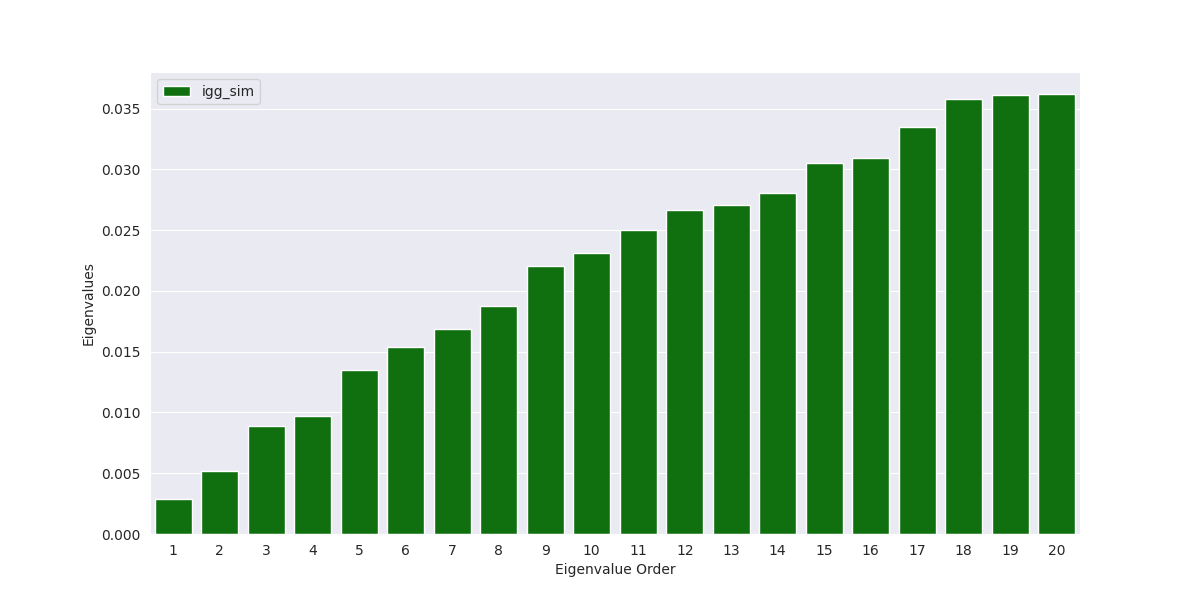}
    \caption{
    {The spectrum of the eigen-decomposition of the Diffusion Maps Laplacian matrix $\LL$ computed for the simulated \ig dataset. The smoothness of the spectrum and having only one 0 eigenvalue(not displayed) indicates that $\mathcal{M}_s$ is a smooth connected manifolds.}
    }
    \label{fig:igg_eigvals}
\end{figure*}

\begin{figure*}
    \begin{minipage}{.4\textwidth}
        \centering
        \subfloat[The $\Phi_e$ embedding { of the experimental hemgagglutinin data}.]{\includegraphics[width=\linewidth]{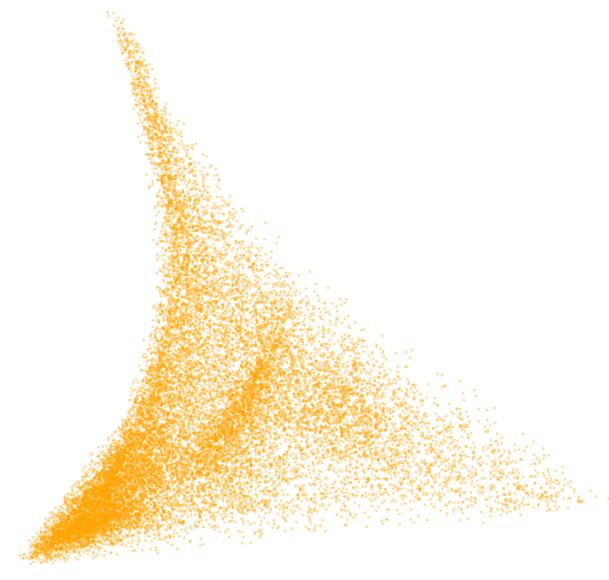}}
    \end{minipage}
    \begin{minipage}{.4\textwidth}
        \centering
        \subfloat[A random sample from $p_e$ embedded into the $\Phi_e$ space using kernel interpolation.]{\includegraphics[width=\linewidth]{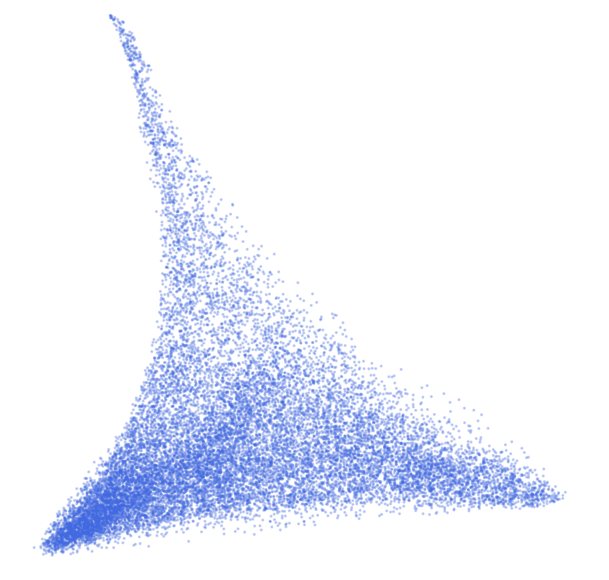}}
    \end{minipage}
    \begin{minipage}{.4\textwidth}
        \centering
        \subfloat[The $\Phi_e$ embedding(orange) and a random sample from $p_e$(blue) embedded into the $\Phi_e$ space using kernel interpolation.]{\includegraphics[width=\linewidth]{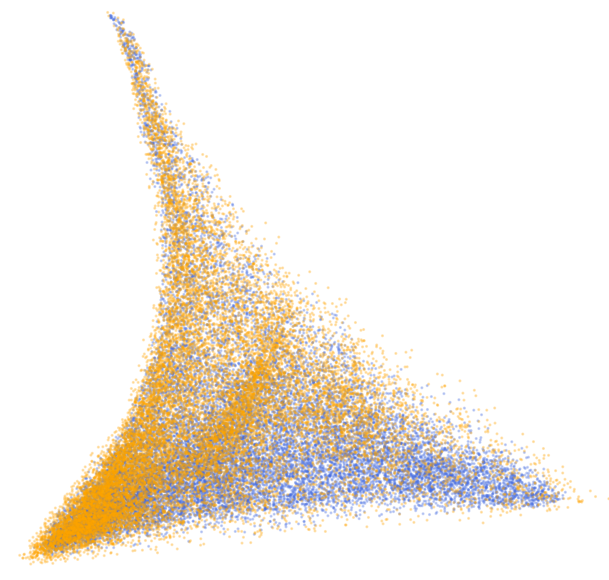}}
    \end{minipage}
    \caption{{ All of the three-dimensional the plots above are rotated to best display the
embedding, with views consistent with Appendix Figure~\ref{fig:exp_embeddings}}. In \textbf{(b)}, we display a random sample from $p_e$, the density on $\mathcal{M}_e$, which we embed into the $\Phi_e$ space \textbf{(a)} using the kernel interpolation method presented in Appendix \ref{sup_sec:interpolation}. We observe that this sample has no gaps and no clusters. In \textbf{(c)}, we display the difference in density between the sample from $p_e$ (blue) and the sample used to compute $\Phi_e$ (orange). This is due to the resampling method described in Appendix \ref{app:uniform_resampling} that aims to mimic a uniform distribution over $\mathcal{M}_e$. We note that $p_e$ is much denser in the low SNR regions (see Figure \ref{fig:exp_embeddings}). By sampling less from this noisy and uninformative region, we encourage $\Phi_e$(orange) to better capture the geometry of $\mathcal{M}_e$.}
    \label{fig:exp_density}
\end{figure*}

 \begin{figure*}
     \centering
  \includegraphics[width=\textwidth]{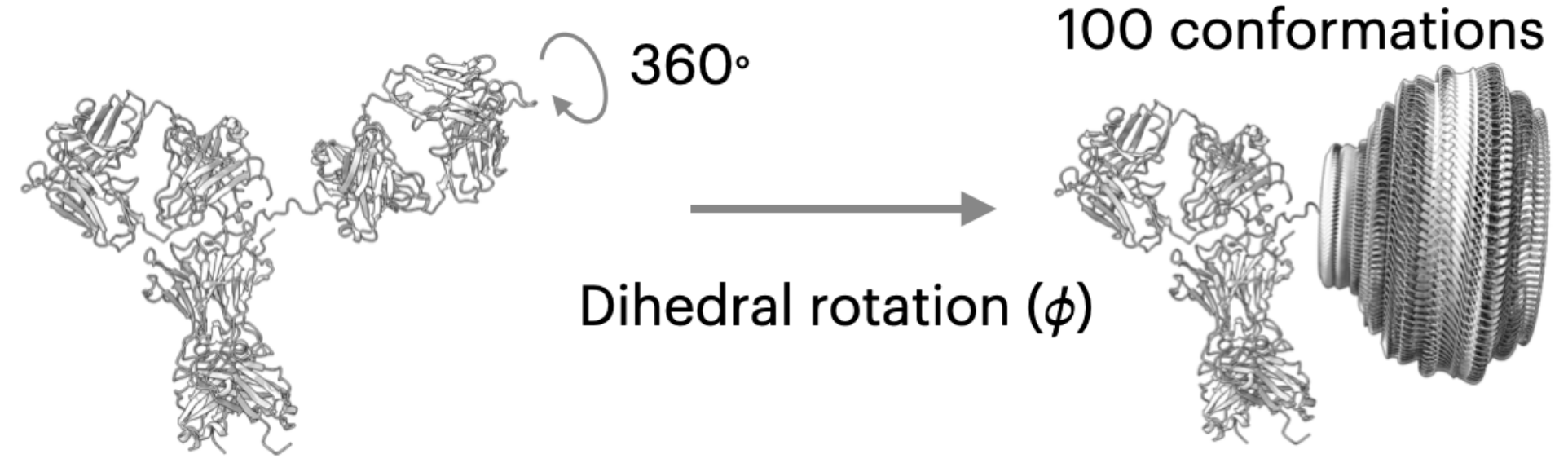} 
     \caption{{ Figure from~\cite{jeon2024cryobench}. Visualization of the conformational heterogeneity of the IgG complex by rotation of a dihedral angle.}
 } 
     \label{fig:igg_atomic_models}
 \end{figure*}

\begin{figure*}[h]
\begin{minipage}{.8\textwidth}
        \centering
        \subfloat[ { Representative structures from normal mode analysis on hemagglutinin structure (PDB id: 6wxb)}]
        {\includegraphics[width=\linewidth]{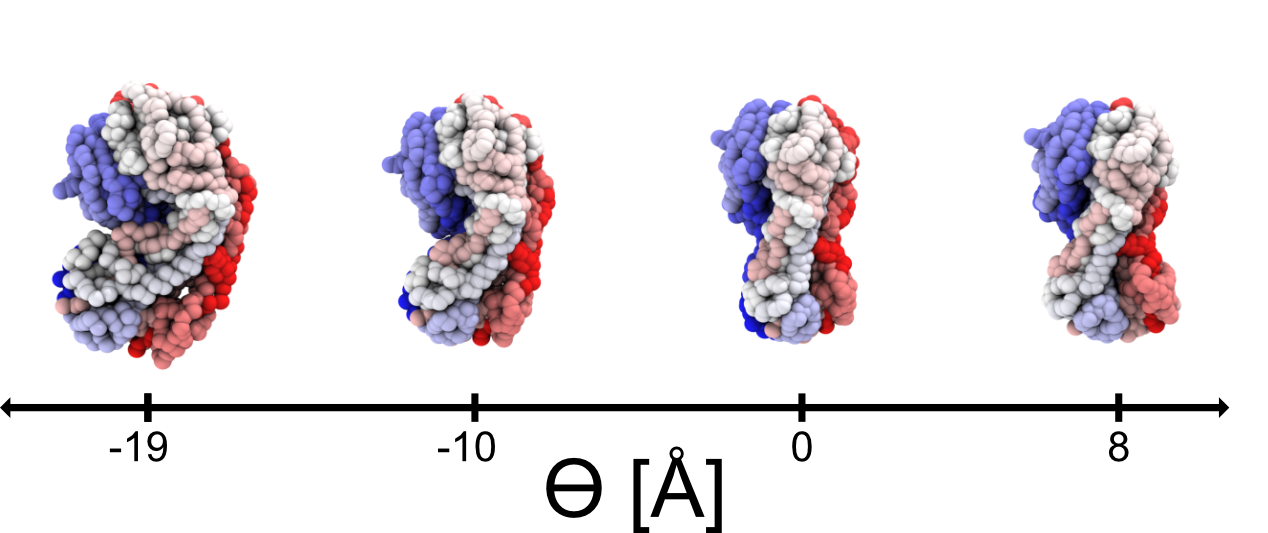}}
    \end{minipage}
    \begin{minipage}{.8\textwidth}
        \centering
        \subfloat[{color{red} Representative non-whitened images from EMPIAR 10532, with whitened images}]
        {\includegraphics[width=\linewidth]{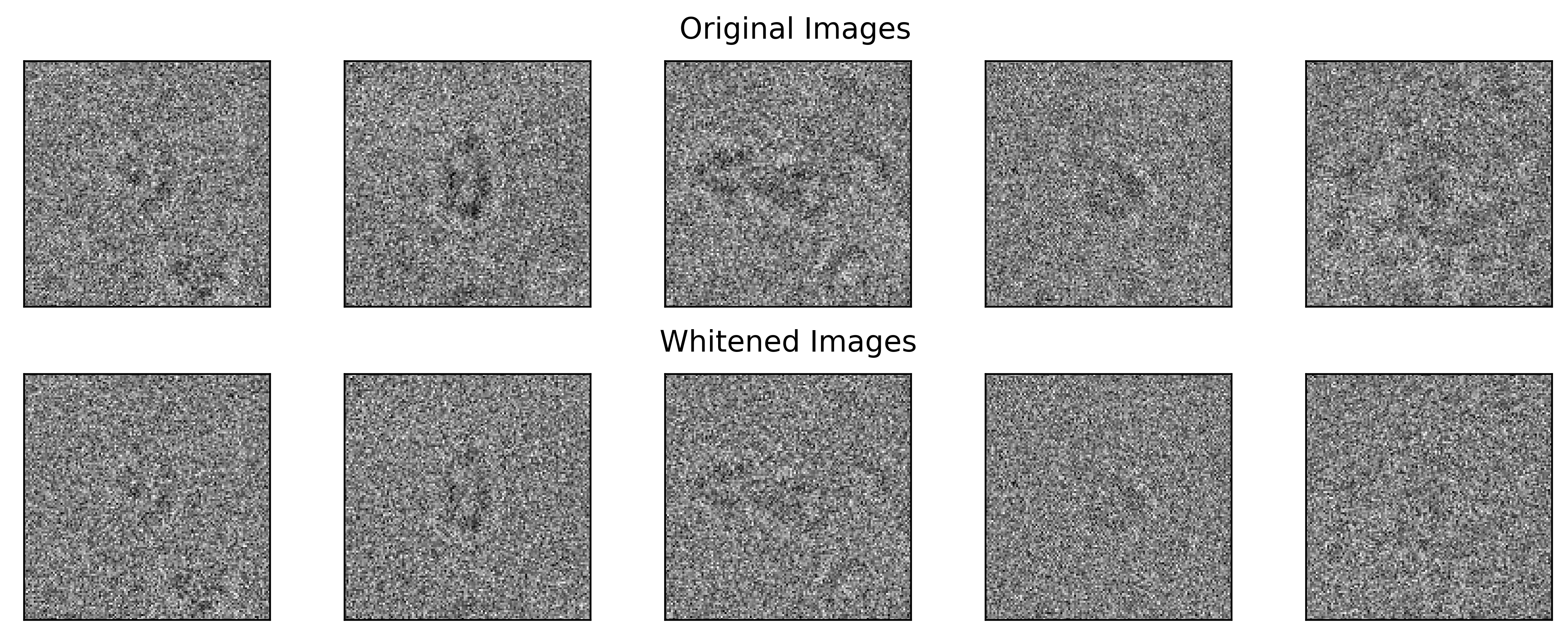}}
    \end{minipage}
\caption{{ Visualization of structures and images used for the hemagglutinin examples, following analysis from~\cite{dingeldein2025amortized}. In a), the structures are obtained from concatenating two normal modes, as detailed in~\cite{dingeldein2025amortized}. The magnitude $|\theta|$ denotes the RMSD from the ground truth structure, with $\theta<0$ for the first normal mode chosen, and $\theta > 0$ for the second chosen normal mode. In b), on the top row we show representative images from EMPIAR 10532~\cite{tan2020through}, and on the bottom row show the images after whitening.}} 
\label{fig:hemagglutinin}
\end{figure*}

\begin{figure*}[ht]
    \centering
    \begin{minipage}{.60\linewidth}
    \includegraphics[width=\linewidth]{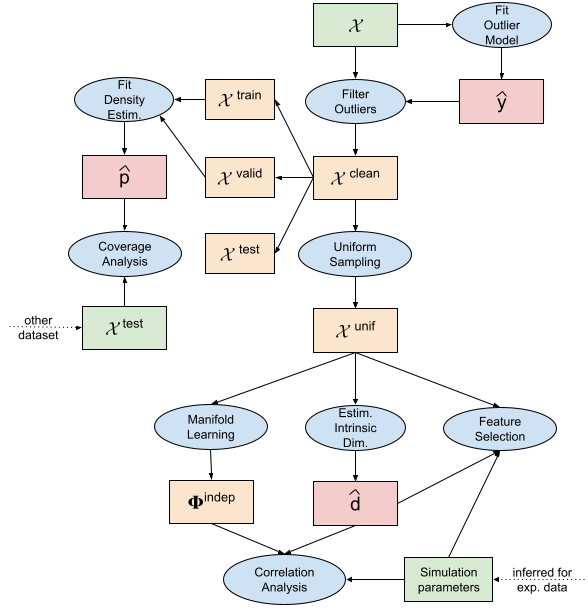}
    \end{minipage}
    \caption{
    {Proposed pipeline for validating the embedding space learned by the encoder $S_\psi$.}
    }
    \label{fig:framework}
\end{figure*}

\begin{figure*}[ht]
    \centering
    \includegraphics[width=0.5\linewidth]{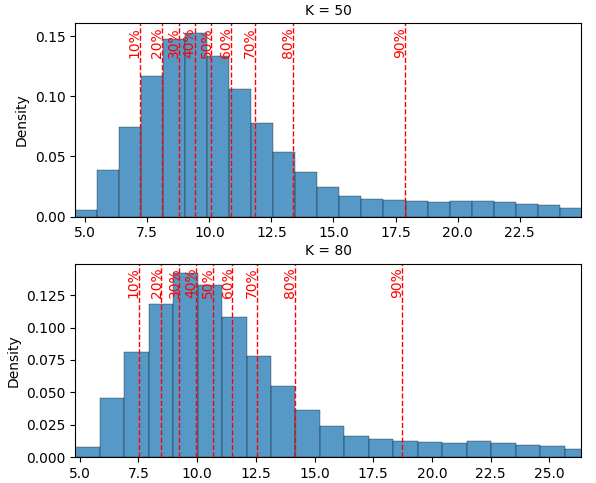}
    \caption{
    { Illustration of the MVS threshold selection heuristic. We show empirical histograms of the distribution of $k$-nearest neighbor distances for two values $k=50$ and $k=80$ for $\X_s \sim p_s$ collected from the \ig dataset. Vertical dashed lines indicate deciles. Each histogram typically has a long right tail corresponding to sparse, low-density outliers. A good choice of $\alpha$ removes this tail, producing a roughly unimodal, bell-shaped distribution; In our \ig dataset, this occurs around $\alpha \approx 0.2$. 
    }
    }
    \label{fig:alpha_selection}
\end{figure*}

\begin{figure*}
    \begin{minipage}{.8\textwidth}
        \centering
        \subfloat[$\Phi_e$ before(left) and after(right) Riemannian Relaxation.]{
            \includegraphics[width=0.5\textwidth]{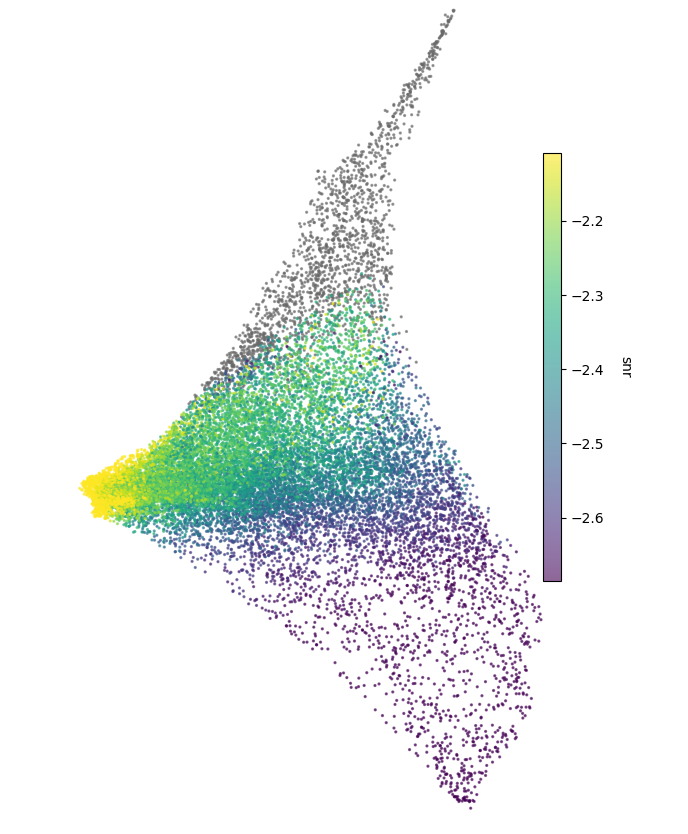}
            \includegraphics[width=0.5\textwidth]{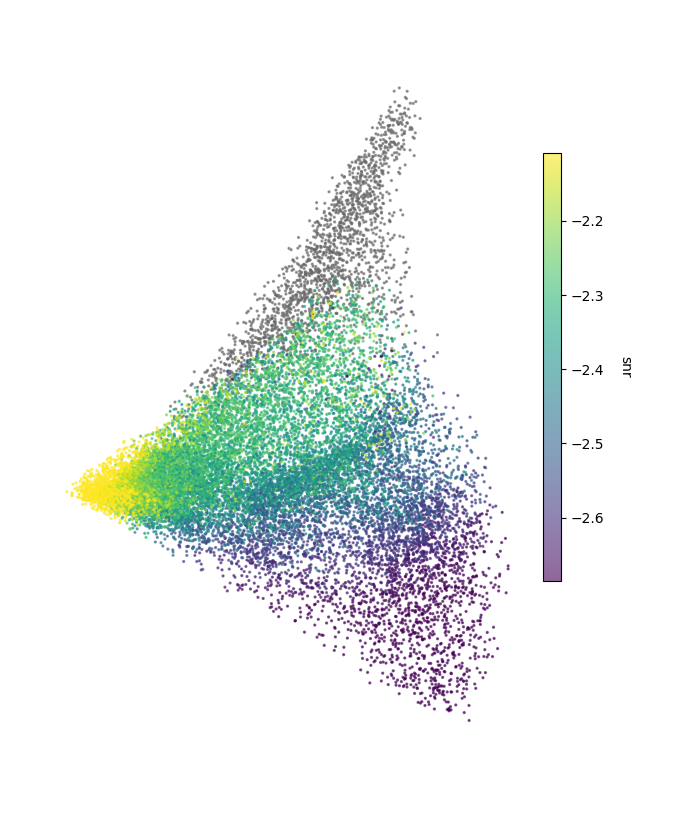}
        }
    \end{minipage}
    \begin{minipage}{.8\textwidth}
        \centering
        \subfloat[$\Phi_s$ before(left) and after(right) Riemannian Relaxation.]{
            \includegraphics[width=0.5\textwidth]{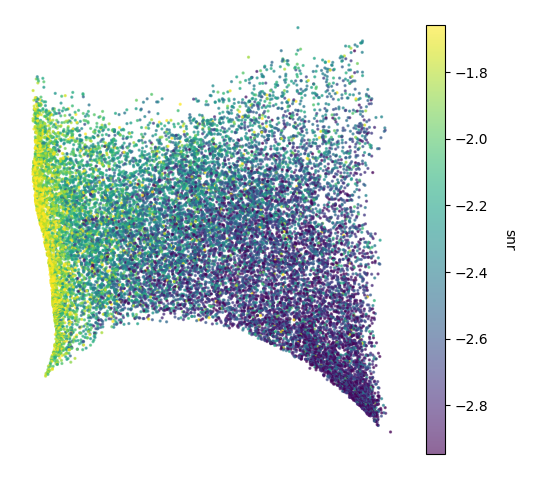}
            \includegraphics[width=0.5\textwidth]{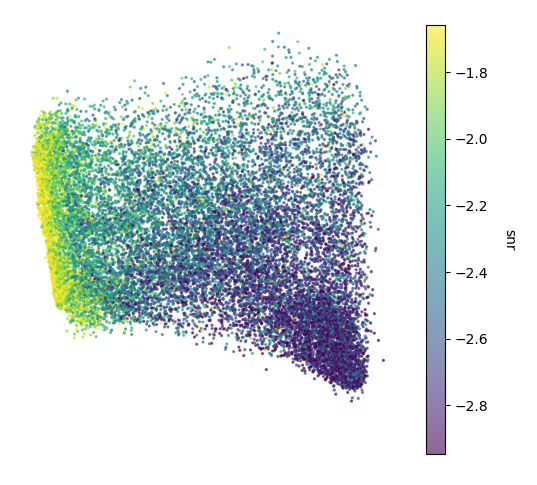}
        }
    \end{minipage}
    \caption{{ Three-dimensional} diffusion maps embeddings $\Phi_e$ \textbf{(a)} and $\Phi_s$ \textbf{(b)} { of the hemagglutinin data} before and after Riemannian Relaxation. The embeddings have been slightly rotated to emphasize the effect of the relaxation. Riemannian Relaxation tends to produce smoother embeddings with less curvature and more uniformly distributed points.}
    \label{fig:relaxation}
\end{figure*}

    

\begin{figure*}
    \centering
    \subfloat[TSLasso results for the \hem experimental data.]{
        \includegraphics[width=0.8\textwidth]{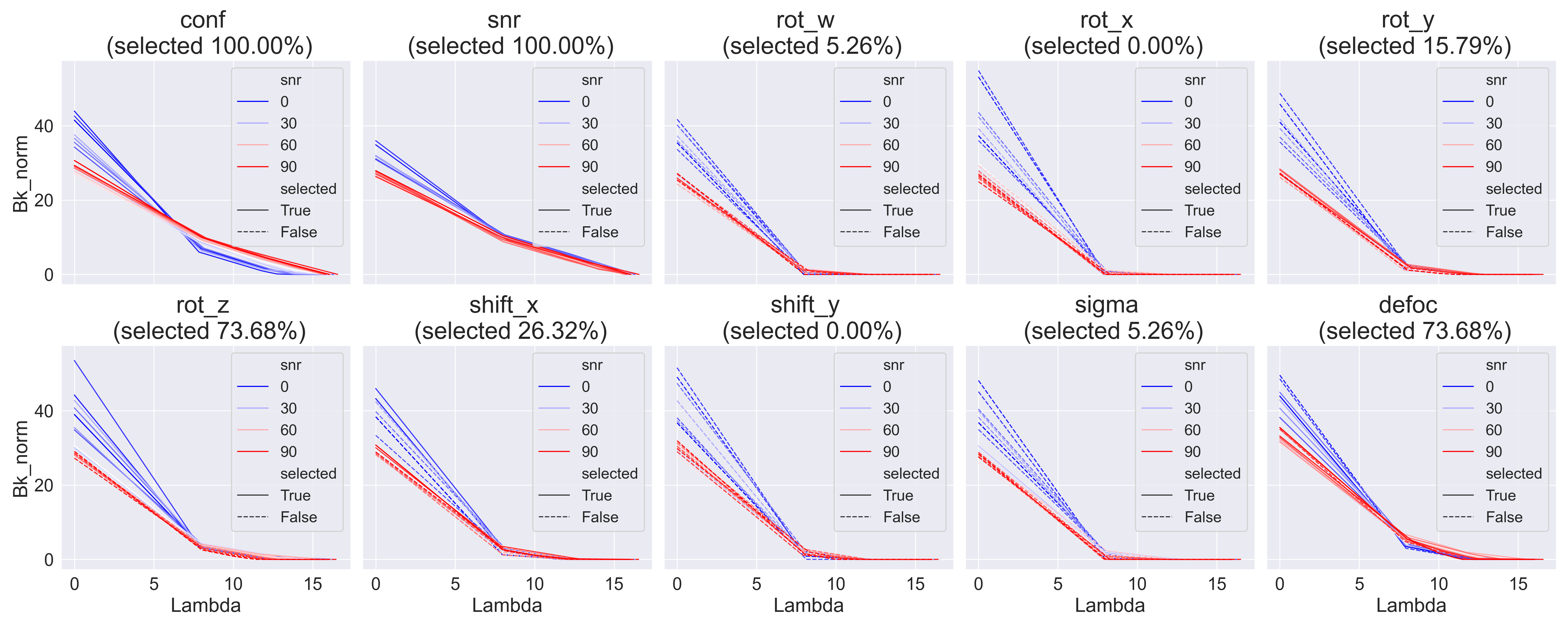}
    }
    \vspace{0.01cm}
    \subfloat[TSLasso results for the \hem simulated data.]{
        \includegraphics[width=0.8\textwidth]{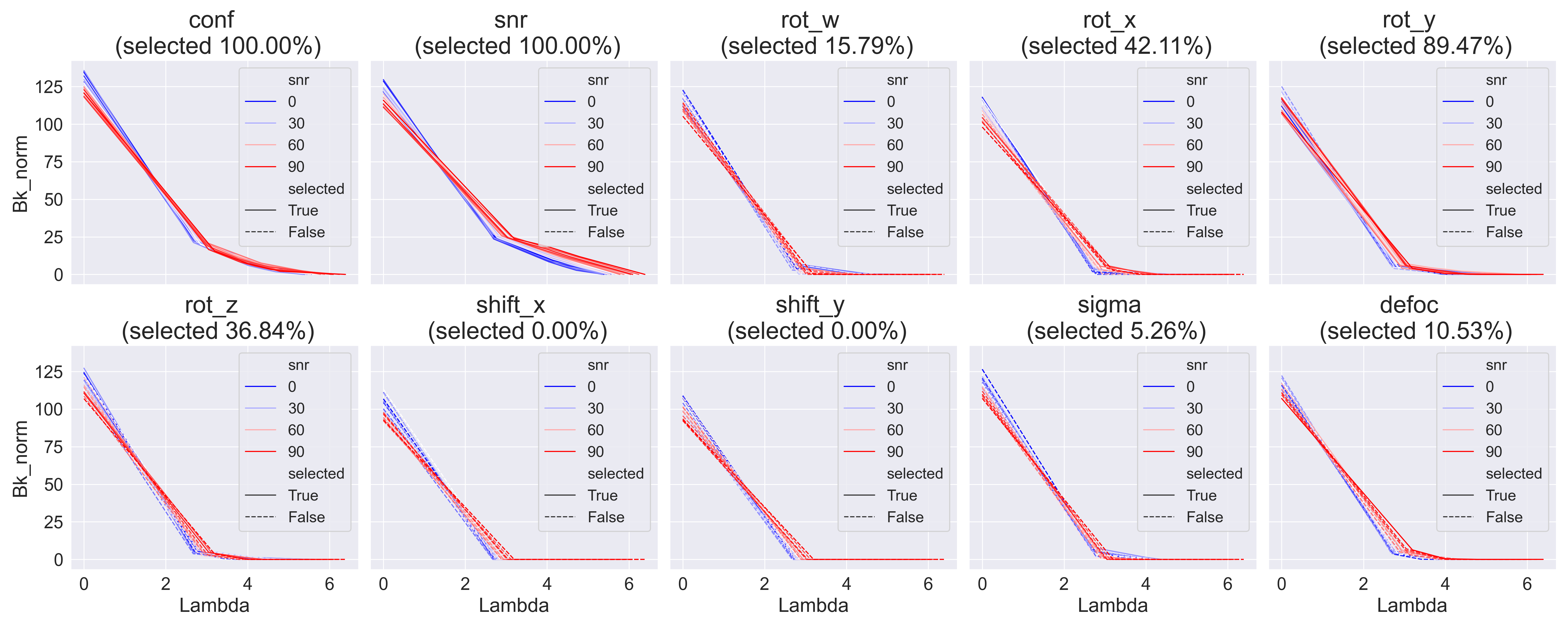}
    }
    \vspace{0.01cm}
    \subfloat[TSLasso results for the \ig simulated data.]{
        \includegraphics[width=0.8\textwidth]{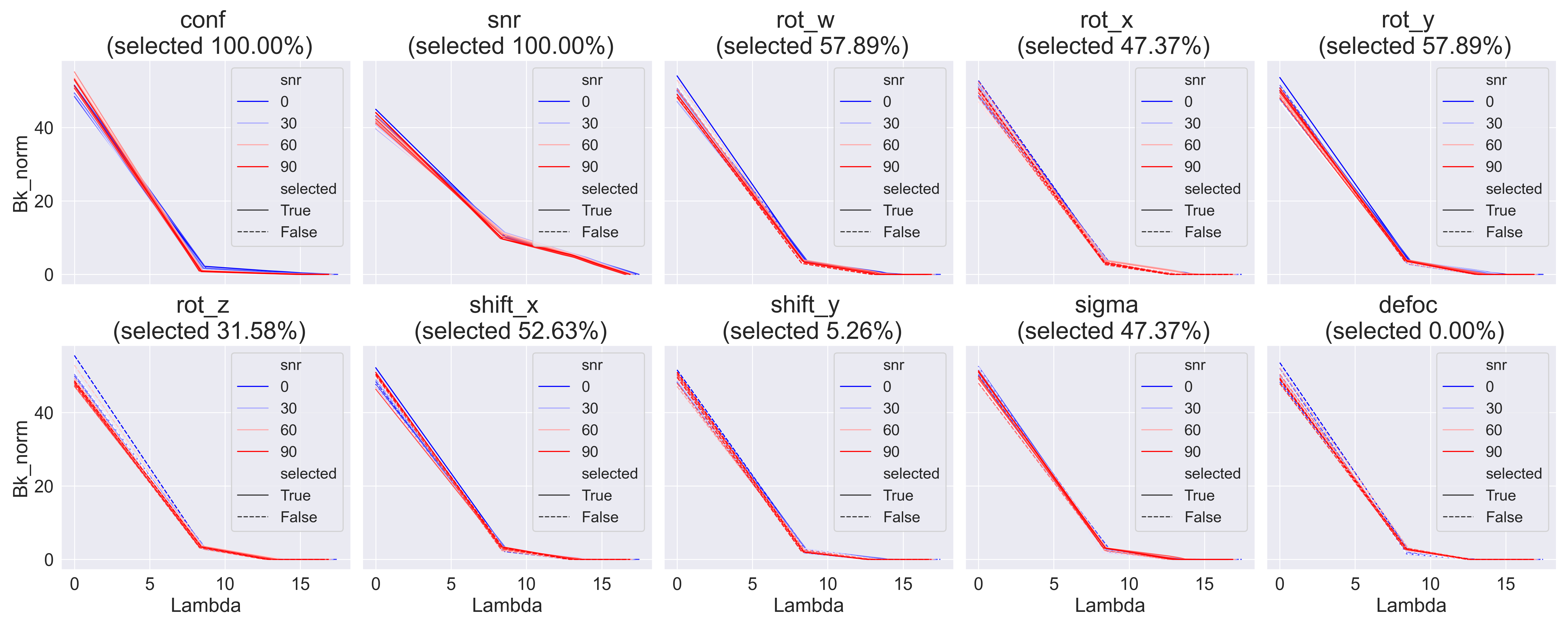}
    }
    \caption{
    {
    The regularization paths of each $f_k \in \mathcal{F}$ obtained over 20 runs of TSLasso for the experimental \textbf{(a)} and simulated \textbf{(b)} \hem data, as well as for the simulated \ig data \textbf{(c)}. Each subplot corresponds to one function $f_k \in \mathcal{F}$, with the name and the selection rate in $f_S$, the solution dictionary identified by TSLasso being indicated in the sub-title. These selection rates should be compared with the chance of picking $d$ functions out of $s$ available ones at random which is $40\%$($d=4$) for the \hem dataset and $50\%$($d=5$) for the \ig dataset, in both cases the dictionary comprising of $s = 10$ candidates. The x-axis represents the value of $\lambda$, the strength of the sparsity regularization, while the y-axis represents the average magnitude of $B_k$, the linear coefficients.  Each run consists only of points in the top $q$-th percentile over all points in terms of $\mathrm{SNR}$. We perform the experiment for $q \in \{0, 5, \dots, 90\}$ with the lines going from blue to red as $q$ increases. A continuous (dotted) line indicates that $f_k$ was selected (not selected) in that run. We find that $f_S$ almost always consists of conformation $\theta$(inferred $\widetilde{\theta}$ for the experimental data), $\mathrm{SNR}$(or inferred $\widetilde{\mathrm{SNR}}$ for the experimental data).}
    }
    \label{fig:tslasso}
\end{figure*}

\end{document}